\documentclass[12pt,a4paper]{article}

\usepackage{epsfig}
\usepackage{psfig}
\def\lsim{\raise0.3ex\hbox{$<$\kern-0.75em\raise-1.1ex\hbox{$\sim$}}}
\def\gsim{\raise0.3ex\hbox{$>$\kern-0.75em\raise-1.1ex\hbox{$\sim$}}}
\makeatletter
\@addtoreset{equation}{section}
\makeatother

\renewcommand{\arraystretch}{1.25}
\newcommand{\beqn} {\begin{equation}}
\newcommand{\eqn} {\end{equation}}
\newcommand{\eqa}{\begin{eqnarray}}
\newcommand{\ena}{\end{eqnarray}}

\newcommand{\slsh}[1] {#1\kern-.43em/}

\newcommand{\real}{{\sf I}\kern-.12em{\sf R}}
\newcommand{\comp}{{\sf I}\kern-.48em{\sf C}}

\newcommand{\nin} {\in\kern-.6em/}

\newcommand{\plaq}{\mbox{\raisebox{-0.40mm}
{\epsfig{bbllx=211,bblly=332,bburx=360,bbury=481,
file=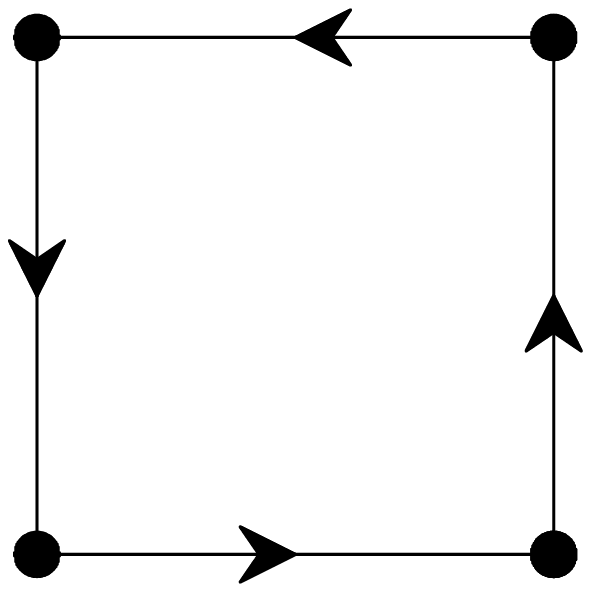,height=4mm}}~}}
\newcommand{\loopt}{\mbox{\raisebox{-2.7mm}
{\epsfig{bbllx=137,bblly=258,bburx=435,bbury=556,
file=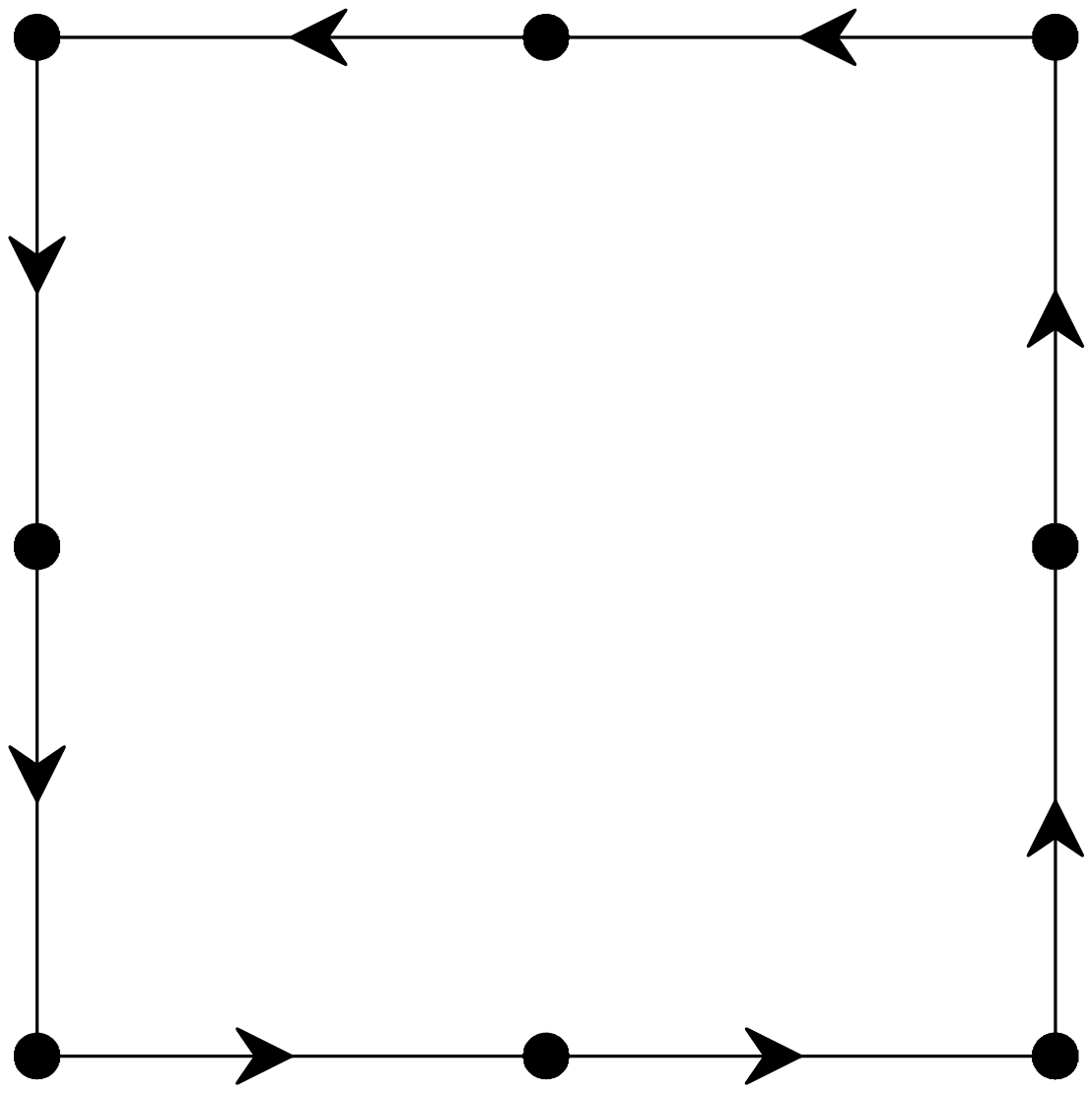,height=8mm}}~}}
\newcommand{\loOp}{\mbox{\raisebox{-0.5mm}
{\epsfig{bbllx=137,bblly=332,bburx=435,bbury=481,
file=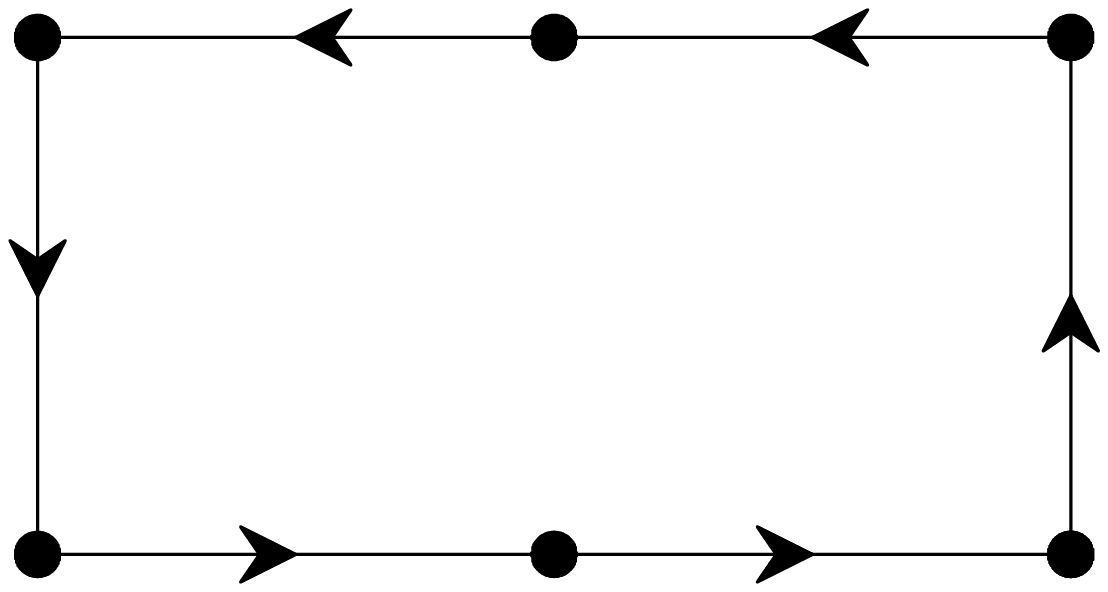,height=4mm}}~}}
\newcommand{\lOop}{\mbox{\raisebox{-2.7mm}
{\epsfig{bbllx=211,bblly=258,bburx=360,bbury=556,
file=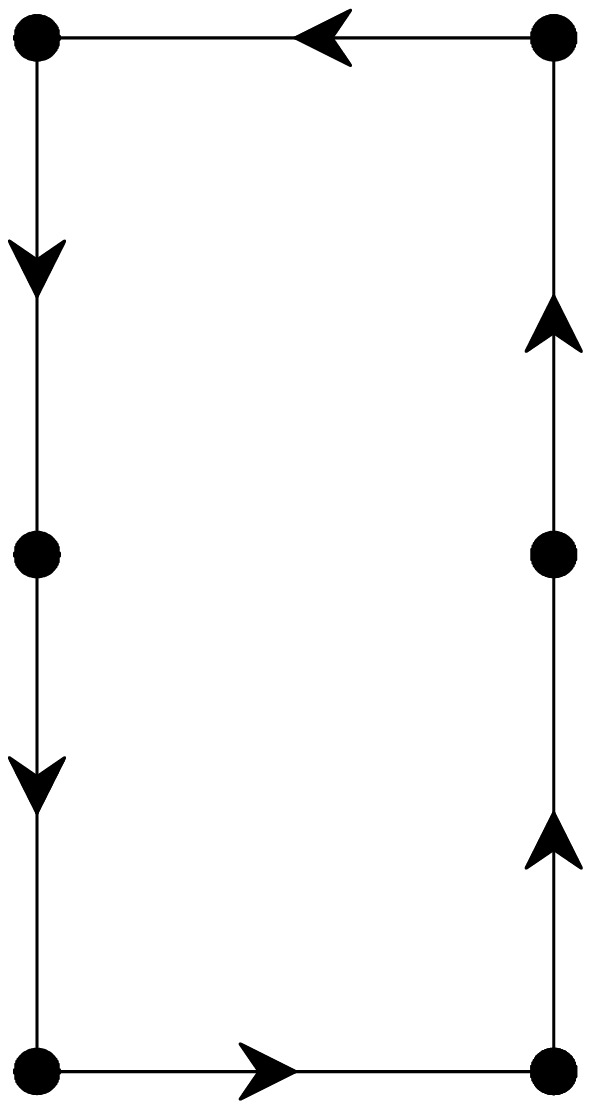,height=8mm}}~}}
\newcommand{\re}{{\rm Re\,}}
\newcommand{\tr}{{\rm Tr\,}}
\newcommand{\nn}{\nonumber}

\def\MEF{m_{\rm eff}}\def\mef{\ifmmode\MEF\else$\MEF$\fi}

\def\NT{N_\tau}
\def\nt{\ifmmode\NT\else$\NT$\fi}
\def\NS{N_\sigma}
\def\ns{\ifmmode\NS\else$\NS$\fi}

\def\PR{{ Phys.\ Rev.\ }}

\def\PRL{{ Phys.\ Rev.\ Lett.\ }}
\def\PL{{ Phys.\ Lett.\ }}
\def\NP{{ Nucl.\ Phys.\ }}

\def\p{^\prime}

\def\n{\noindent}
\def\nn{\nonumber}
\def\bn{\bigskip\noindent}

\begin{document}
\thispagestyle{empty}

 \mbox{} \hfill BI-TP 99/02\\
 \mbox{} \hfill April 1999
\begin{center}
\vspace*{1.0cm}
{{\large \bf Determination of Anisotropy Coefficients\\
              for SU(3) Gauge Actions from the\\
                Integral and Matching Methods\\}
 }
\vspace*{1.0cm}
{\large J. Engels$^{\rm a}$, F. Karsch$^{\rm a}$ and 
T. Scheideler$^{\rm b,}$\footnote[1]
{Present address: IBM, P.O. Box 265, 00101 Helsinki, Finland.} } \\
\vspace*{1.0cm}
{\normalsize
$\mbox{}$ {$^{\rm a}$Fakult\"at f\"ur Physik, Universit\"at Bielefeld,
D-33615 Bielefeld, Germany}\\
\bigskip
$\mbox{}$ {$^{\rm b}$Department of Physics, P.O.Box 9, 
00014 University of Helsinki, Finland}}\\
\vspace*{2cm}
{\large \bf Abstract}
\end{center}
\setlength{\baselineskip}{1.3\baselineskip}

We use two non-perturbative methods to obtain the anisotropy
derivatives of the coupling constants (the anisotropy coefficients)
of $SU(3)$ lattice gauge theory. These coefficients appear in the derivative
formulae for the energy density and the pressure. We calculate them for 
the standard Wilson and two improved actions, the $(2\times 2)$ and the
Square Symanzik action.   
Both methods lead for all investigated actions to compatible
results which are clearly different from their known asymptotic 
perturbative limits. With increasing $\beta$ the limits are however
approached in all cases. Our findings strongly support the equivalence   
of the integral and derivative methods for the calculation of energy density 
and pressure.

\newpage

\setcounter{page}{1}


\section{Introduction}

For the investigation of the phase transition from ordinary hadronic
matter to the quark gluon plasma and the approach of the plasma to its
high temperature limit the calculation of the energy density $\epsilon$
and the pressure $p$ is of central importance. These observables are defined 
as derivatives of the partition function $Z$ with respect to temperature $T$ 
and volume $V$
\beqn 
\epsilon = - {1 \over V} {\partial \ln Z \over \partial(1/T)}~,\quad\quad
   p        = T {\partial \ln Z \over \partial V}~. 
\label{press}
\eqn
\n In the lattice formulation of QCD the temperature and the volume 
therefore have
to be independent variables. On a lattice with a fixed number
of points \ns~ in each spatial direction, and \nt~ points in the temporal
direction, this is achieved by choosing corresponding independent lattice
spacings $a_{\sigma}$ and $a_{\tau}$ so that 
\beqn
 V =(\ns a_{\sigma})^3\quad{\rm and }\quad T^{-1} = \nt a_{\tau}~.
\eqn
\n As a consequence two different coupling 
constants have to be introduced for the spatial and temporal contributions 
to the action. For $SU(N_c)$ lattice gauge theory we consider actions 
of the following form
\beqn
S = \beta_{\sigma} \sum_{x,\mu < \nu <4} S_{\mu \nu}(x)
    + \beta_{\tau} \sum_{x,\mu  <4} S_{\mu 4}(x)~,
\eqn
\n where $S_{\mu \nu}(x)$ is a generalized plaquette in the $(\mu,\nu)-$plane.
Generalized plaquette expectation values are then defined by
\eqa
P_{\sigma} &=& {1 \over 3\NS^3 \NT} \big < \sum_{x,\mu < \nu <4} 
               S_{\mu \nu}(x) \big> \\ 
P_{\tau}   &=& {1 \over 3\NS^3 \NT} \langle \sum_{x,\mu <4} 
               S_{\mu 4}(x) \rangle~.
\ena
\n The couplings $\beta_{\sigma,\tau}$ may be rewritten in terms of the 
coupling constants $g^2_{\sigma,\tau}$ as
\beqn
 \beta_{\sigma}={2N_c\over g^2_{\sigma}}\cdot{1\over \xi}\quad{\rm and}\quad 
   \beta_{\tau}  ={2N_c\over g^2_{\tau }}\cdot \xi~.
\eqn
\n Here $\xi=a_{\sigma}/a_{\tau}$ is the {\it anisotropy}. One  
may also express the couplings with the {\it bare anisotropy} $\gamma$ 
and a common coupling $\beta$ through \cite{burg}
\beqn
\beta = \sqrt{\beta_{\sigma} \beta_{\tau}}={2N_c\over g^2}~,\quad 
\gamma = \sqrt{{\beta_{\tau} \over \beta_{\sigma}}} 
       = \xi\cdot{g_{\sigma}\over g_{\tau}}~,
\eqn
or equivalently
\beqn
\beta_{\sigma}={\beta\over \gamma}~,\quad  \beta_{\tau} = \beta \gamma~.
\eqn
\n The couplings depend on $a_{\sigma}$ and $a_{\tau}$, or, alternatively,
on $a_{\sigma}$ and $\xi$. Following ref. \cite{class} we adopt the 
latter two as independent variables. The energy density and pressure are
then
\beqn
\epsilon = -{T \over V} \Big< \xi {\partial S \over \partial \xi }\Big>~,
\quad \quad 
\epsilon -3p = {T \over V} \Big< a_{\sigma} {\partial S \over \partial 
a_{\sigma} }\Big>~.  
\eqn
Obviously, the energy density contains now terms, which are proportional
to derivatives with respect to the anisotropy $\xi$, because
\eqa
\xi {\partial \beta_{\sigma} \over \partial \xi} \!&=&\!
-\beta_{\sigma}+2N_c{\partial g^{-2}_{\sigma} \over \partial \xi}~,\\
\xi {\partial \beta_{\tau} \over \partial \xi} \!&=&\!
\beta_{\tau}+2N_c \xi^2{\partial g^{-2}_{\tau} \over \partial \xi}~.
\ena
After taking the derivatives one can evaluate the expressions on isotropic
lattices. At $\xi=1$ we have
\beqn
\gamma=1~,\quad g_{\sigma}=g_{\tau}=g~, \quad a_{\sigma}=a_{\tau}=a~,
\label{iso}
\eqn
and the relation for the $\beta-$function \cite{Frit82}
\beqn
a{dg^{-2} \over da} = -2 \left ({\partial g^{-2}_{\sigma} \over \partial \xi}
+{\partial g^{-2}_{\tau} \over \partial \xi} \right)_{\xi=1} ~. 
\label{bfunc}
\eqn
Energy density and pressure are now conveniently combined in the two equations
\eqa
{\epsilon +p \over T^4}\!\! &=&\!\! 8N_c\nt^4 g^{-2}\left[ 1-{g^2 \over 2} \left(
{\partial g^{-2}_{\sigma} \over \partial \xi}-
{\partial g^{-2}_{\tau} \over \partial \xi}\right)_{\xi=1}\right]
(P_{\sigma}-P_{\tau})~,\label{epp}\\
{\epsilon -3p \over T^4}\!\! &=&\!\! 12N_c\nt^4
\left ({\partial g^{-2}_{\sigma} \over \partial \xi}
+{\partial g^{-2}_{\tau} \over \partial \xi} \right)_{\xi=1}
[2P_0-(P_{\sigma}+P_{\tau})]~.
\label{emdp}
\ena
Here $P_0$ denotes a generalized plaquette on a symmetric ($T=0$) lattice. 
Its inclusion normalizes both $\epsilon$ and $p$ to zero at $T=0$.  

We define the anisotropy coefficients $c_{\sigma,\tau}$ as follows
\beqn
c_{\sigma,\tau}(a) \equiv \left({\partial g^{-2}_{\sigma,\tau} \over
 \partial \xi} \right)_{\xi=1}~.
\eqn
In the weak coupling limit, they are related to the derivatives of the
anisotropic parts of the couplings
\beqn
g^{-2}_{\sigma,\tau}(a,\xi) = g^{-2}(a) + c^w_{\sigma,\tau}(\xi) +O(g^2)~,
\eqn
by
\beqn
c_{\sigma,\tau}(0) = \left({\partial c^w_{\sigma,\tau}(\xi) \over
\partial \xi} \right)_{\xi=1}.
\eqn
The quantities $c^w_{\sigma,\tau}(\xi)$ were calculated in ref.\cite{Frit82}
for the Wilson action perturbatively. For the difference one obtains
\beqn
c_{\sigma}(0)-c_{\tau}(0) = {N_c^2-1 \over N_c} \cdot 0.146711-N_c\cdot 
0.019228~,
\label{cdi}
\eqn
and, from the renormalization group equation 
and eq. (\ref{bfunc})  
\beqn
c_{\sigma}(0)+c_{\tau}(0) = b_0 = {11N_c \over 48 \pi^2}~.
\label{csu}
\eqn
These perturbatively calculated anisotropy coefficients were used in
early finite temperature lattice calculations. In $SU(3)$ gauge theory,
however, this led to unphysical behaviour of the pressure: it could 
become negative and at the critical point a gap in $p$ was observed
\cite{Svet83,Deng89}. A second, rather important consequence of 
taking only the perturbative coefficients is the rapid approach of
$\epsilon$ and $p$ to their respective ideal gas limits soon after the
transition, though the limits themselves are independent of 
$c_{\sigma,\tau}$. The actual form of the approach is of course dependent 
on the residual interaction in the quark gluon plasma, and therefore of 
physical relevance.

In ref. \cite {Eng90} an alternative method for the calculation of energy
density and pressure was proposed, which is now in common use.
Here the pressure is obtained without the anisotropy
coefficients from an integral over plaquettes. In order
to be consistent, the integral method should be equivalent to the
use of non-perturbative anisotropy coefficients, i.e. the corresponding
values for $c_{\sigma,\tau}$ should be essentially independent on finite
size and cut-off effects. First results in ref.\cite{Boyd96} showed 
however considerable cut-off effects for the Wilson action. To clarify
the situation, we started \cite{Eng98} a non-perturbative calculation of the 
coefficients with the matching method, which we want to complete in this 
paper.

In the following section we discuss the integral method in some more detail 
as in \cite{Boyd96} and \cite{Eng98} and apply it to
the Wilson action. The matching method is described and applied to the 
same action in section 3. For comparison we determine in section 4 
the anisotropy coefficients as well for two improved actions, the 
$(2\times 2)$ and the Square Symanzik action \cite{Garca}.
We close with a summary and the conclusions.
 

\section{The integral method}

For large homogeneous systems with isotropic interactions, the volume
derivative in the pressure formula (\ref{press}) may be replaced by a volume 
division. The pressure and the free energy density are then related by
\beqn
p = -f = {T \over V}\ln Z~.
\label{pef}
\eqn 
Instead of calculating ln$Z$ directly, one may take advantage of the relation
\beqn
{\partial \ln Z \over \partial \beta} = - \Big< {\partial S \over \partial
\beta} \Big> ~,
\eqn
and integrate the measured expectation value over a $\beta-$range. For
$\xi=1$ we obtain then
\beqn
{p \over T^4} = \left.{p \over T^4}\right|_{\beta_0} + 
\int \limits_{\beta_0}^{\beta} d\beta\p [2P_0-(P_{\sigma}+P_{\tau})]~.
\eqn
As in eqs. (\ref{epp}) and (\ref{emdp}), the inclusion of the $P_0-$term 
serves to normalize 
$p$ at $T=0$ to zero. Obviously, the last equation will lead to a continuous
pressure, even if there is a gap in the plaquettes at the critical point.
The energy density is found from 
\beqn 
{\epsilon -3p \over T^4} = -6N_c\nt^4
a{dg^{-2} \over da} [2P_0-(P_{\sigma}+P_{\tau})]~,
\eqn
by just adding $3p/T^4$. Here we have replaced the sum of the coefficients by
the $\beta-$function. This function and/or $a=a(g^2)$ have to be known 
non-perturbatively, not only to evaluate the last equation, but also to 
determine the temperature $T=1/\nt a$ on fixed $\nt-$lattices at any given 
coupling. Parametrizations of the 
$\beta-$func\-tion for the $SU(3)$ Wilson action have been derived by several 
groups \cite{Boyd96,Akem93,Scri97}. They are rather similar above $\beta=5.9$,
but below this value they differ from each other. Probably this is due to the 
transition from the weak coupling to the strong coupling regime.  

\subsection{The high temperature limit}

The two methods for the calculation of $\epsilon$ and $p$ on the lattice 
suffer from cut-off effects, especially for small $\nt-$values. Though these
cut-off effects are of similar size, there is a non-negligible difference.
For the same method, the high temperature limits of $\epsilon$ and $p$ have
the same cut-off dependencies, because in both cases the same formula for 
$(\epsilon -3p)/T^4$ is used and 
\beqn 
(\epsilon -3p)/T^4 = O(g^4)~. 
\eqn
\begin{table}
\begin{center}
\begin{tabular}{|r|c|c|c|}
\hline
$\nt$ & $R_D$ & $R_I$ & $r=R_D/R_I$  \\
\hline
    4  & 1.4952 & 1.3778 & 1.0852  \\
    6  & 1.1816 & 1.1323 & 1.0435  \\
    8  & 1.0867 & 1.0659 & 1.0195  \\
\hline
\end{tabular}
\end{center}
\caption{The ratios $R(\nt)=\epsilon(\nt)/\epsilon_{SB}=p(\nt)/p_{SB}$ for the
Wilson action calculated with the derivative ($R_D$) and the 
integral methods ($R_I$).}
\label{tab:ratios}
\end{table}
\n However, as different methods are based on the evaluation of different
operators, which have different cut-off dependencies, the high$-T$ limit
derived in different approaches leads also to distinct cut-off dependencies.
In the derivative method the high temperature limit is derived from the lowest 
order of the $g^2-$expansion of the energy density ($\xi=1$)
\eqa
{\epsilon \over T^4} &=& 6N_c\nt^4 g^{-2} (P_{\sigma}-P_{\tau}) +O(g^2)~,\\
  &=& {3\over 2} (N_c^2-1)\nt^4 (P_{\sigma}^{(2)}-P_{\tau}^{(2)})+O(g^2)~.
\ena
The quantities $P_{\sigma,\tau}^{(2)}$ are the corresponding 
expansion coefficients
of the generalized plaquette expectation values
\beqn
P_{\sigma,\tau} = g^2{N_c^2-1 \over 4N_c} P_{\sigma,\tau}^{(2)}+O(g^2)~.
\eqn
For the Wilson action one obtains the following deviation 
of the energy density, $\epsilon(\nt)$, in the high $T$ limit
from the continuum result, $\epsilon_{SB}$ \cite{Bein96}
\beqn
R_D = {\epsilon(\nt) \over \epsilon_{SB}} = 1 +{10 \over 21}\left(
{\pi \over \nt}\right)^2 +{2 \over 5}\left({\pi \over \nt} \right)^4 + 
O\left({1 \over \nt^6}\right)~.
\label{RSW}
\eqn 
If the integral method is used, the calculation of the high temperature limit
starts from eq.(\ref{pef}) and the lowest order term of ln$Z$. Here the $T=0$   
subtraction has to be done explicitly. With the Wilson action one arrives at 
\beqn
R_I = {f(\nt) \over f_{SB}} = 1 +{8 \over 21}\left({\pi \over \nt}
\right)^2 +{5 \over 21}\left({\pi \over \nt} \right)^4 + 
O\left({1 \over \nt^6}\right)~,
\label{RIW}
\eqn
where of course $\epsilon_{SB}=-3f_{SB}$. The full $1/\nt^2-$expansions
can be calculated numerically. In Table~\ref{tab:ratios} we list them for
the two methods in the case of the Wilson action and $\nt=4,6,8\,$. We show
also the ratio $r$ between the two high temperature limits.
We stress that this difference, which expresses itself in the ratio $r$,
will play a crucial role in the discussion of the anisotropy coefficients 
presented in the following sections.

\subsection{Determination of the anisotropy coefficients}

\n
As mentioned already in the introduction, the integral method implies the
use of non-perturbative anisotropy coefficients. We can determine them by 
comparing the pressure formulae from the two different approaches
\beqn
\left({p \over T^4} \right)_D = r \cdot\left({p \over T^4} \right)_I~,
\label{preseq}
\eqn
and by exploiting the $\beta-$function relation, eq.(\ref{bfunc}). In the
pressure equation (\ref{preseq}) we have inserted the factor $r$ as a 
compensation for the differing cut-off effects. In principle, $r=r(g^2)\,$.
We approximate $r$ by its value at $g^2=0$ (Table \ref{tab:ratios}).
From eqs.(\ref{epp}) and (\ref{emdp}) we obtain
\beqn
 \left({p \over T^4} \right)_D = N_c\nt^4 \left[ \left( 2g^{-2}-
(c_{\sigma}-c_{\tau}) \right) (P_{\sigma}-P_{\tau}) -3(c_{\sigma}+c_{\tau})
\Bigl( 2P_0-(P_{\sigma}+P_{\tau}) \Bigr) \right]~,
\eqn
and therefore
\eqa
\!\!c_{\sigma}\!\!\!&=&\!\!\!g^{-2} - {a\over 4}{d g^{-2} \over da}\nonumber\\
          \!\!\!& &\!\!\!+{1 \over\ P_{\sigma}-P_{\tau}} \left[
 - {r \over 2N_c\nt^4} \left({p \over T^4} \right)_I 
 + {3a\over 4}{d g^{-2} \over da}\Bigl( 2P_0-(P_{\sigma}+P_{\tau}) \Bigr) 
 \right]~,\label{csi}\\
\!\!c_{\tau}\!\!\!&=&\!\!\!-c_{\sigma} -{a\over 2}{d g^{-2} \over da}~.
\ena

\begin{figure}[hp]
\begin{center}
   \epsfig{bbllx=94,bblly=264,bburx=483,bbury=587,
       file=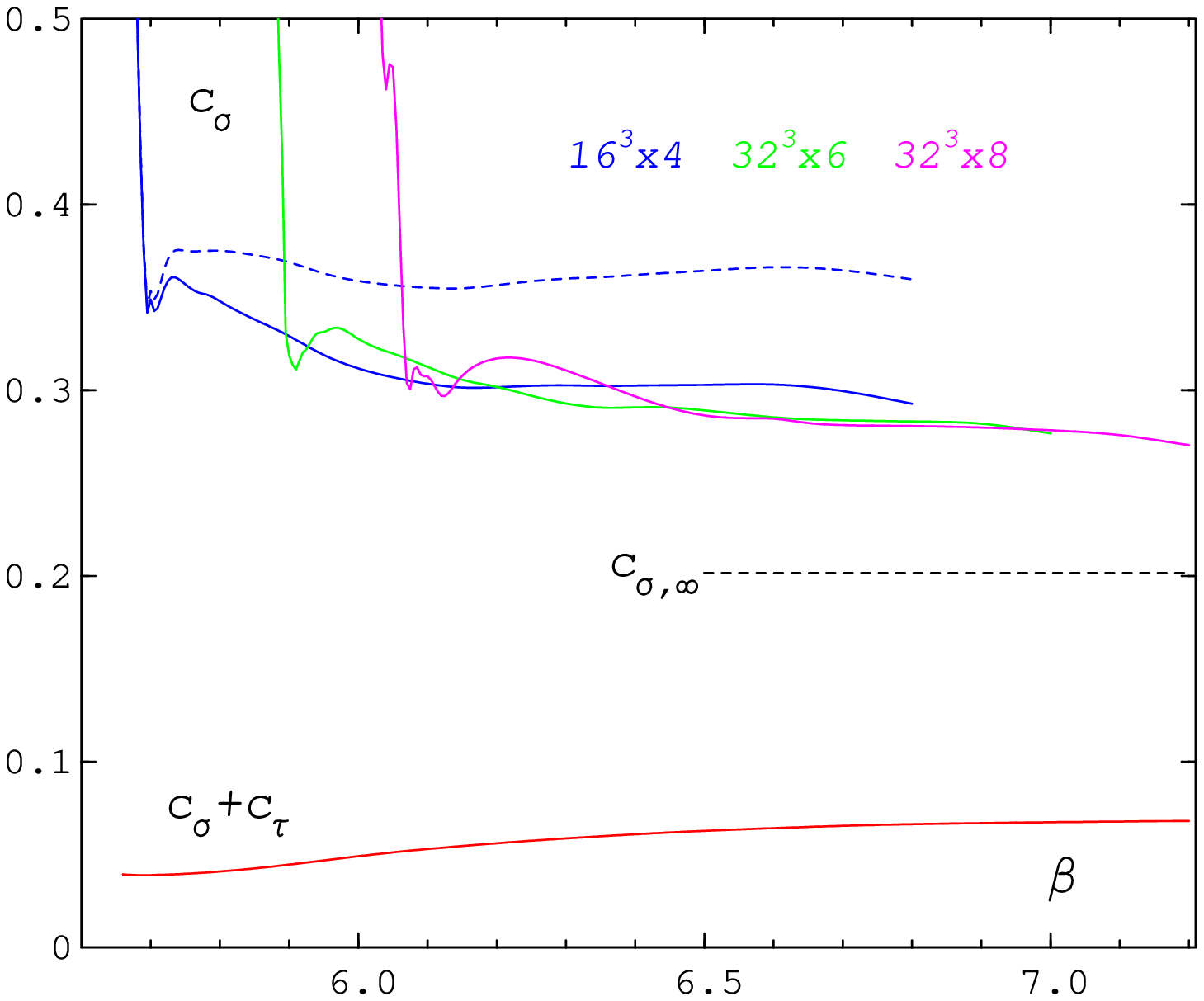, width=100mm,height=83mm}
\end{center}
\caption{The coefficient $c_{\sigma}$ for the $SU(3)$ Wilson action from 
plaquette interpolations. The dashed curve is without cut-off corrections
from ref. \cite{Eng98}. Also shown is the perturbative result for $\beta
\rightarrow \infty\,$, $c_{\sigma,\infty}$ and the sum $c_{\sigma}+c_{\tau}$.} 

\label{fig:cs3a}
\begin{center}
   \epsfig{bbllx=94,bblly=264,bburx=483,bbury=587,
       file=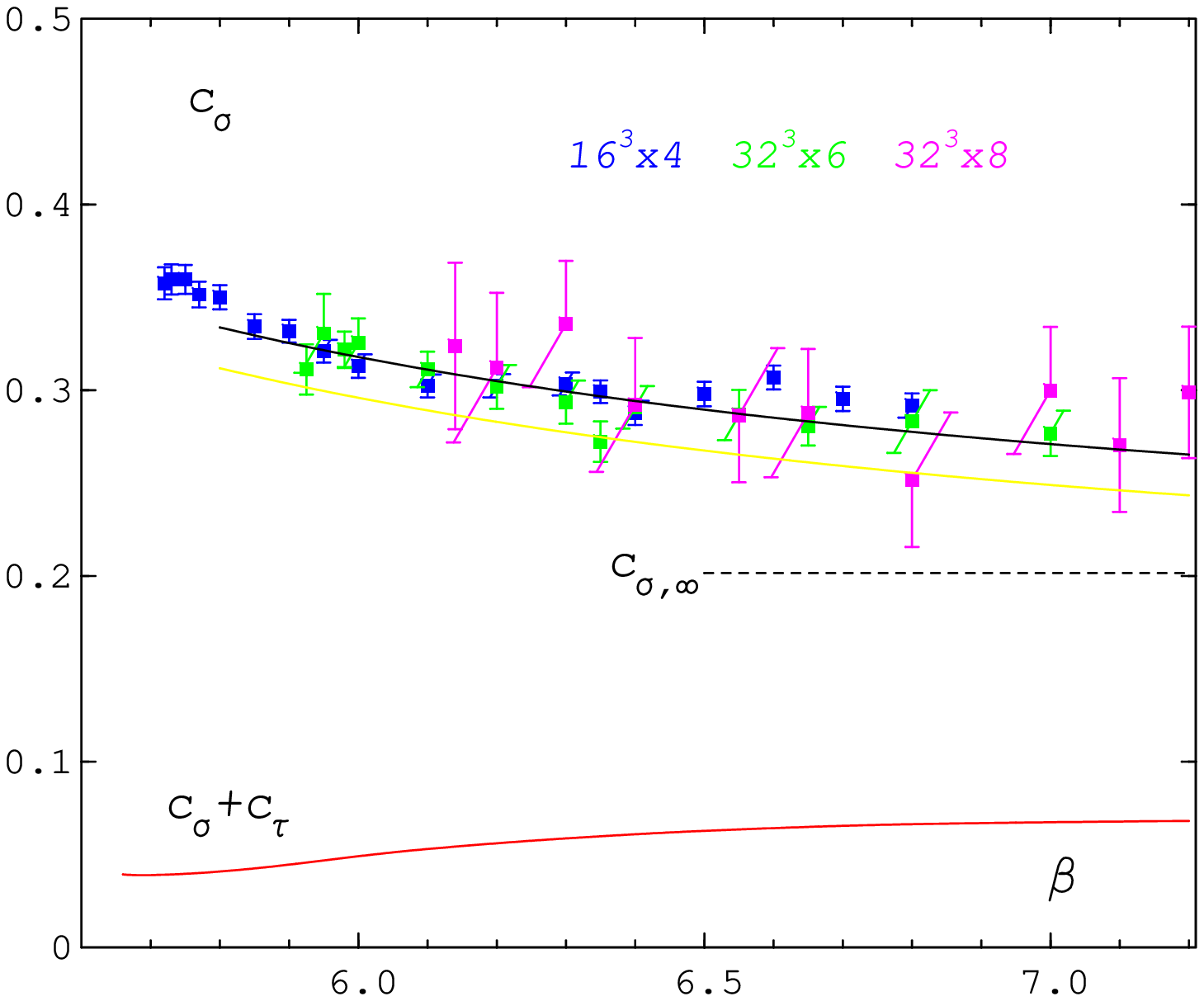, width=100mm,height=83mm}
\end{center}
\caption{The coefficient $c_{\sigma}$ for the $SU(3)$ Wilson action 
with errors coming from the plaquette data . The solid curve is the fit, 
eq. (\ref{pade}), the yellow 
line a corresponding fit to data (not shown), which one obtains if
$p/T^4(\beta_0=5.65,\nt=4)=0.05$~.} 

\label{fig:cs3b}
\end{figure}

In the following we discuss the behaviour of $c_{\sigma}$. The behaviour of
$c_{\tau}$ is similar, because both coeffcients are related by the smooth
$\beta-$function.
It is clear, that the numerical determination of the coefficients becomes 
problematic below the transition point, because there 
both the pressure and the plaquette differences are small.
Indeed, this is seen in Fig. \ref{fig:cs3a}, where we show $c_{\sigma}$ for the
$SU(3)$ Wilson action as obtained from eq. (\ref{csi}) on $\nt=4,6,8-$lattices
using interpolations of the $[P_{\sigma}-P_{\tau}]$ 
and $[P_0-(P_{\sigma}+P_{\tau})/2]-$data of ref. \cite{Boyd96}.
Apart from the close vicinity of the critical

\newpage
\n  $\beta-$values the results
for different $\nt$ are consistent with each other. For comparison we show 
also the previous result for $\nt=4$ from ref. \cite{Eng98}, where the 
difference in the cut-off effects of the two methods was not taken into account, 
that is $r=1$ was used. Obviously the influence of the $r-$factor is important.
Moreover it leads to slightly lower results for $c_{\sigma}$, though,
as before, there remains a substantial deviation from the asymptotic value  
$c_{\sigma,\infty}\equiv c_{\sigma}(0)$ from eqs. (\ref{cdi}) and (\ref{csu}).

It is difficult to estimate all possible error sources in the calculation of 
$c_{\sigma}$, as for  example that of the parametrization of the 
$\beta-$function. The major errors are certainly due to the errors in the 
plaquette expectation values. In Fig. \ref{fig:cs3b} we present 
 the results including these errors, but omitting the 
respective critical regions.
We can summarize these data with a simple Pad\'e fit of the form
\beqn
c_{\sigma} = c_{\sigma,\infty} {1+d_1 g^2+d_2 g^4 \over 1+d_0 g^2}~,
\label{pade}
\eqn
where $d_0=-0.64907 ,~d_1=-0.61630 ,~d_2=0.16965$~.

A further source of
errors is the result for $(p/T^4)_I$~. Whereas the integration itself
contributes only a negligible error, the size of the pressure at the lower 
integration boundary $\beta_0$ is unknown. We have identified 
$\beta_0$ with the point where the integrand $[P_0-(P_{\sigma}+P_{\tau})/2]$
is disappearing and assumed, as usual, that there $p/T^4$ is zero. This 
may be not true. In order to assess the effect of a non-zero pressure
contribution we have repeated our calculation with a test value  
$p/T^4=0.05$ at $\beta_0=5.65,T_0/T_c=0.917$ for $\nt=4$. 
As a result the value of $c_{\sigma}$
is diminished by 0.022 in the whole $\beta-$range considered. The same
shift in $c_{\sigma}$ is observed also for $\nt=6$ and 8. For comparison 
we have plotted in Fig. \ref{fig:cs3b} a fit of the form (\ref{pade}) to the
shifted data as well. Here we find $d_0=-0.60983 ,~d_1=-0.78889 ,~d_2=0.36164$~.

Some insight into the question of the pressure size below the transition is
gained from the behaviours of the plaquette differences around $T_c$ in
$SU(3)$ and $SU(2)$. In 
Fig. \ref{fig:plaq} we compare them to each other. 
The main difference between the two gauge groups is evidently arising
from the fact, that the respective phase transitions are of different order.
Also, there is no doubt, that in $SU(2)$ the pressure is non-zero well
below $T_c$. 

As we have seen in $SU(3)$ the inclusion of the cut-off factor $r$ 
considerably lowers the $\nt=4-$result for $c_{\sigma}\,$. In $SU(2)$
the values which were previously \cite{Eng95} determined


\setlength{\unitlength}{1cm}
\begin{picture}(13,6.5)
\put(0.2,0){
   \epsfig{bbllx=127,bblly=264,bburx=451,bbury=587,
       file=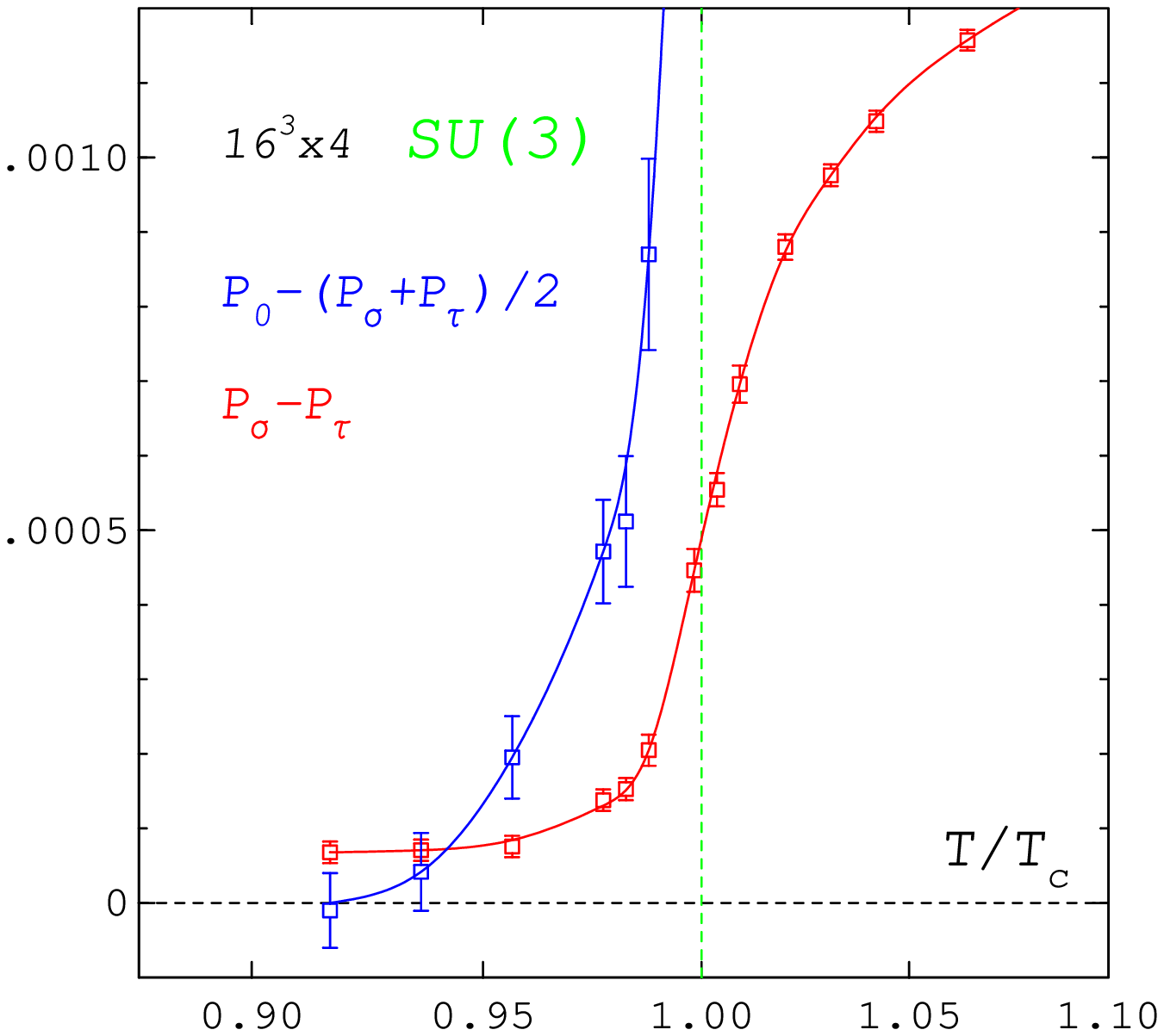, width=60mm,height=60mm}
          }
\put(7.2,0){
   \epsfig{bbllx=127,bblly=264,bburx=451,bbury=587,
       file=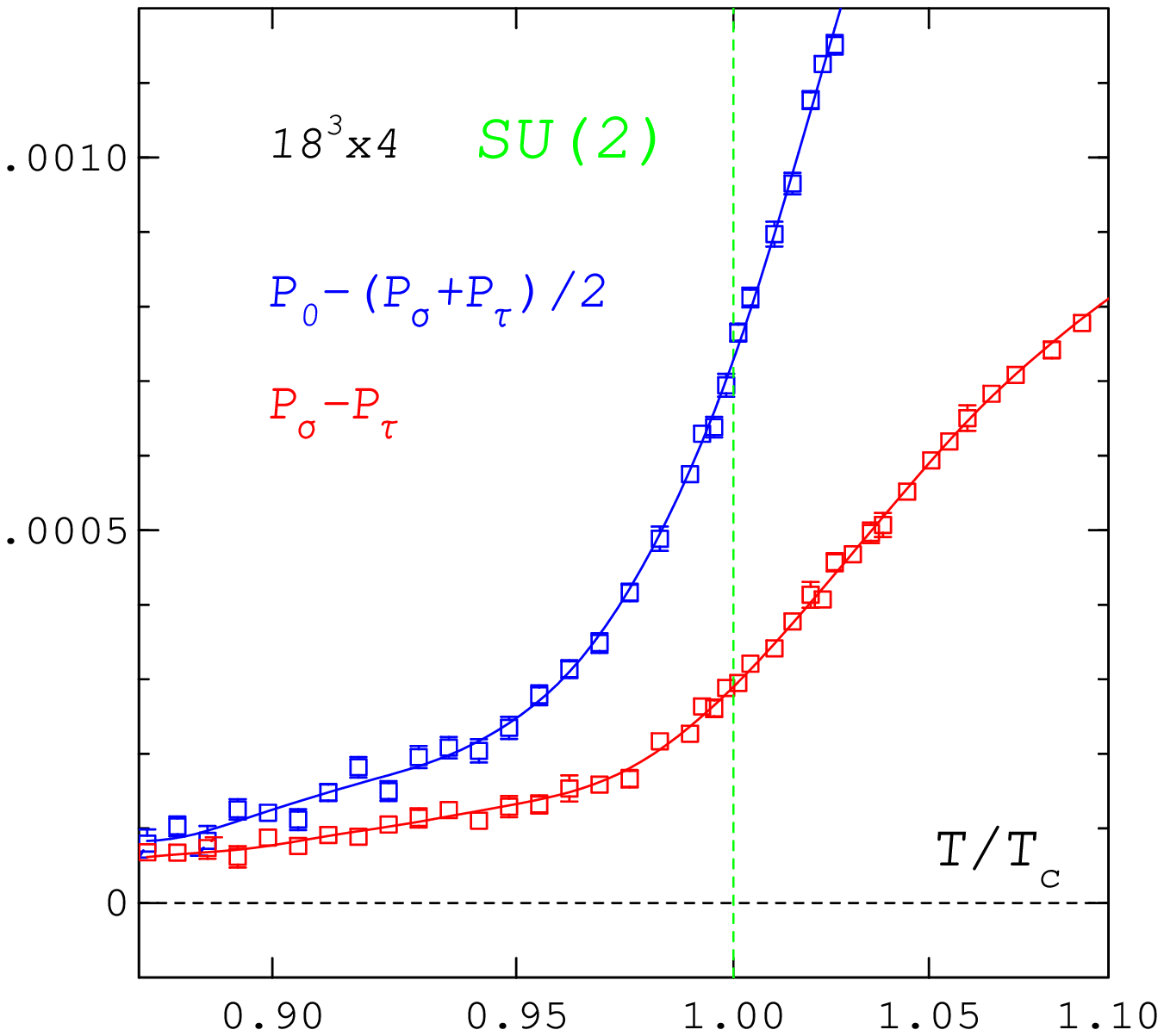, width=60mm,height=60mm}
          }
\end{picture}
\begin{figure}[h]
\caption{The plaquette differences in the vicinity of the deconfinement
transition for $SU(2)$ and $SU(3)$ on $\nt=4 -$lattices. The $SU(2)-$data
are from ref. \cite{Eng99}.}
\label{fig:plaq}
\end{figure}
\begin{figure}[htb]
\begin{center}
   \epsfig{bbllx=94,bblly=264,bburx=483,bbury=587,
       file=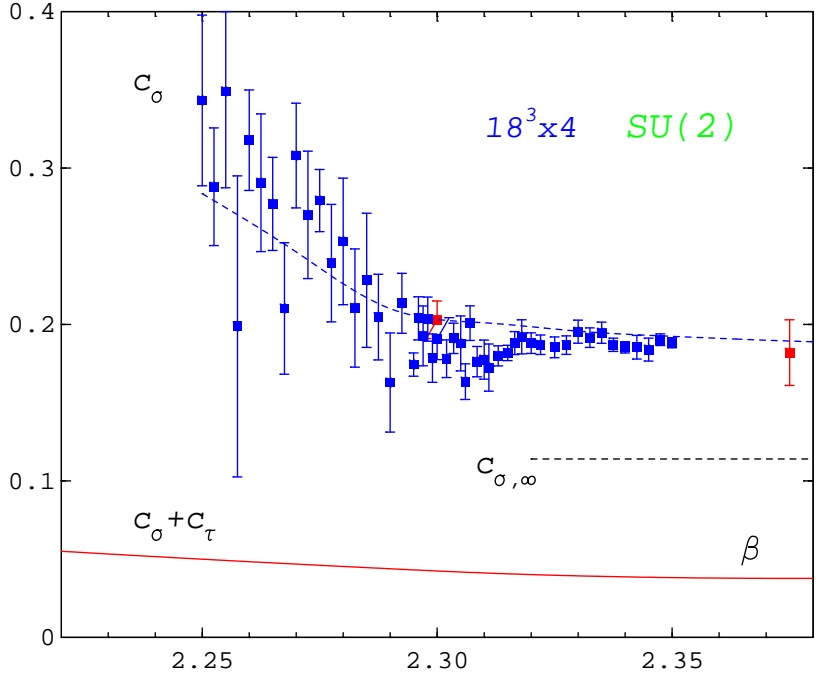, width=100mm,height=83mm}
\end{center}
\caption{The coefficient $c_{\sigma}$ for the $SU(2)$ Wilson action 
with errors coming from the plaquette data and $r=1.0852$. 
The dashed curve $(r=1)$
and the sum $c_{\sigma}+c_{\tau}$ are from
ref. \cite{Eng95}, the red points from ref. \cite{Ejiri}.  
Also shown is $c_{\sigma,\infty}$ .} 
\label{fig:cs2}
\end{figure}

\n 
for $c_{\sigma}$ with 
the integral method and $r=1$ are well in accord with points measured 
in an independent way at $\beta=2.30$ and $\beta=2.375$ by Ejiri et al.
\cite{Ejiri}.
One has to check then, whether this agreement is lost if the 
correct $r$ is taken into account. We do this by using the new high precision
$SU(2)$ data of ref. \cite{Eng99}, already shown in Fig. \ref{fig:plaq}, 
the $\beta-$function of ref. \cite{Eng95} and the $r-$value from Table 
\ref{tab:ratios}. The resulting  $c_{\sigma}$ is plotted in Fig. \ref{fig:cs2}.
We find only a small change in this $\beta-$range, the result is lowered
at $\beta=2.30$ by 0.0067 and at $\beta=2.35$ by 0.0142~.  
The reason for this is,
that up to $\beta=2.35$ the difference $[P_0-(P_{\sigma}+P_{\tau})/2]$ is
still large and only at higher couplings the pressure term with the factor $r$
is dominating in eq. (\ref{csi})~. 
We expect therefore a stronger decrease at higher $\beta-$values.  
\vskip 0.2truecm
 

\section{The matching method}


This method was first used by Burgers et al. \cite{burg} to measure the
anisotropy $\xi$ as a function of the bare anisotropy $\gamma$ on anisotropic
lattices. With this information one may as well determine the anisotropy 
coefficients. Consider the couplings
\beqn
g_{\sigma}^{-2} = {\beta \over 2N_c}{\xi \over \gamma}\quad,\quad
g_{\tau}^{-2} =  {\beta \over 2N_c }{\gamma \over \xi }~,
\eqn
and their derivatives with respect to the anisotropy $\xi$
\eqa
{\partial g^{-2}_{\sigma} \over \partial \xi}\!\!&=&\!\!
g^{-2}_{\sigma}\left(~~{1 \over \xi}  +{\partial \ln\beta \over \partial \xi}
 -{\partial \ln \gamma \over \partial \xi} \right)~,\\
{\partial g^{-2}_{\tau} \over \partial \xi}\!\!&=&\!\!
g^{-2}_{\tau}\left(  -{1 \over \xi}  +{\partial \ln \beta \over \partial \xi}
 +{\partial \ln \gamma \over \partial \xi} \right)~.
\ena
At $\xi=1$ we may apply eq. (\ref{iso}) to find the difference
\beqn
c_{\sigma}-c_{\tau}= 2g^{-2} \left( 1 - {\partial \gamma \over \partial\xi}
\right)_{\xi=1}~,
\eqn
and, using eq. (\ref{bfunc}), we obtain 
\beqn
c_{\sigma,\tau} = \pm g^{-2} \left( 1 - {\partial \gamma \over \partial\xi}
\right)_{\xi=1} -{a \over 4} {dg^{-2} \over da}~.
\label{cmatch}
\eqn
A measurement of the function $\xi(\gamma)$ in the neighbourhood of 
$\xi=1$ will therefore enable us to calculate the anisotropy coefficients,
once we know the $\beta-$function. In principle, the anisotropy $\xi$
may be determined with two measurements of a physical observable which
depends on a distance, which can be chosen in a spatial or in the temporal
direction.  
At the same physical distance the expectation values have to be the same
and thereby fix the anisotropy. This is the idea of matching.

\subsection{Matching of Wilson loop ratios}

\n
Suitable quantities for the matching process are obtained from Wilson loops
of size $n_1\times n_2$ (in lattice units)$\,$. The Wilson loops are related 
to the heavy quark potential
\beqn
W(x_1,x_2) \sim \exp [ -x_1 V_l(x_2) ]~,
\label{Wilson}
\eqn 
for $x_1 \rightarrow \infty\,$. Here, $x_i=n_i a_i$. The potential $V_l$ 
differs from the continuum potential $V$ by a term $V_0$, 
the self-energy of the heavy quarks, which is dependent on
lattice spacing, but independent of $x_2$ \cite{Stack}. The 
natural way to proceed then is to build ratios of Wilson loops, which
depend only on $V_l(x_2)$ for large $x_1$. We use the following ratios
\eqa
R_1(n_1,n_2) &=& {W(x_1+a_1,x_2) \over W(x_1,x_2)} ~;\\
& &\nonumber\\
R_2(n_1,n_2) &=& {W(x_1+2a_1,x_2)\,W(x_1+a_1,x_2) \over W(x_1,x_2)^2}~.
\ena  
On an anisotropic lattice
there are two different types of loops : space-space ($W_{\sigma\sigma}$)
and space-time ($W_{\sigma\tau}$) Wilson loops. The corresponding ratios
$R_{\sigma}$ and $R_{\tau}$ are measuring the same 
potential $V(x_2)$, if the matching condition for the physical distance
\beqn
x_2=y=t \quad {\rm or} \quad n_t=\xi n_y~,
\eqn  
is fulfilled, apart from an $n-$independent factor 
$k=k_{\sigma\sigma}/k_{\sigma\tau}\,$, which is due to the dependence of
$V_0$ on $a_{\sigma}$ and $\xi$
\beqn
R_{\sigma}(n_x,n_y) = k\cdot R_{\tau}(n_x,n_t=\xi n_y)~.
\eqn
At $\xi=1$ the factor $k$ is of course $1\,$. We do not implement any 
smearing for our Wilson loops, because then space and time links would
have to be smeared by the same amount in physical units. However, in 
order to enhance the accuray of the Wilson loop measurements
we are using the fast link integration technique of de Forcrand 
and Roiesnel \cite{Forc} wherever this is feasible. It cannot be done for 
loops with an $n_i=1$ and at loop edges. 
In applying this algorithm we are saving in addition half of the computer
time by taking advantage of the fact that the necessary link contour integrals  
are real and can as well be obtained from a half contour in the complex plane. 

\begin{figure}[htb]
\begin{center}
   \epsfig{bbllx=94,bblly=234,bburx=483,bbury=557,
        file=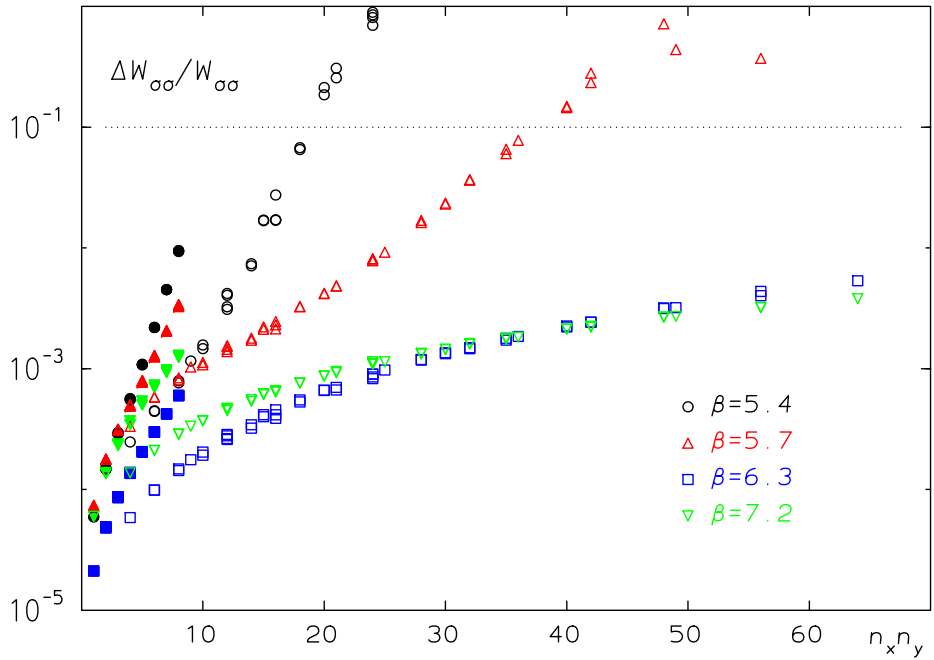,width=84mm,height=70mm,angle=-90}
\end{center}
\caption{The relative errors of spatial Wilson loops as a function of their area
at $\gamma=1.06\,$. Open symbols show the errors of link integrated loops, 
filled symbols those of not integrated ones.} 
\label{fig:err}
\end{figure}

For the Wilson action simulations were performed at four $\beta-$values:
5.4, 5.7, 6.3 and 7.2 using a $16^4$ lattice for $\gamma \in [0.92,1.08]$,
a $16^3\times 32$ lattice for $\gamma \in [1.1,2.0]$ and a $16^3\times 48$ 
lattice for $\gamma =3.0$. The Wilson loops were calculated after every fourth
sweep through the lattice, where a sweep consisted of one heatbath update 
succeeded by four overrelaxation steps. 
The integrated autocorrelation times which one finds then are increasing
with increasing $\beta$ from 0.5, 1.5, 3.0 to over 3.0 for $\beta=7.2\,$.
On the average we took 5000-7000 measurements for $\beta=5.4-6.3$ and 600
for $\beta=7.2\,$. Generally, for the same number of measurements, the 
relative error of a Wilson loop increases with the size of its area. The 
increase itself is much steeper at lower $\beta-$values than at higher ones.
Larger area Wilson loops are therefore only included in the matching procedure
for the higher $\beta-$values. 
In Fig. \ref{fig:err} we show a typical example of this behaviour.
The difference in accuracy between link integrated and plain loops is 
clearly seen in the plot.

After measuring the Wilson loops at a fixed value of $\beta$ and $\gamma$,
we compute the ratios $R_{\sigma}$ and $R_{\tau}\,$. For $\gamma>1$ we connect
the timelike ratios $R_{\tau}$ with spline interpolations. In order to
improve the interpolations we include as well the ratios with $n_t=1\,$.
The spatial ratios $R_{\sigma}$ are then shifted in $n_y$ by a factor $\xi$
and in height 

\begin{figure}[hp]
\begin{center}
   \epsfig{bbllx=94,bblly=264,bburx=483,bbury=587,
       file=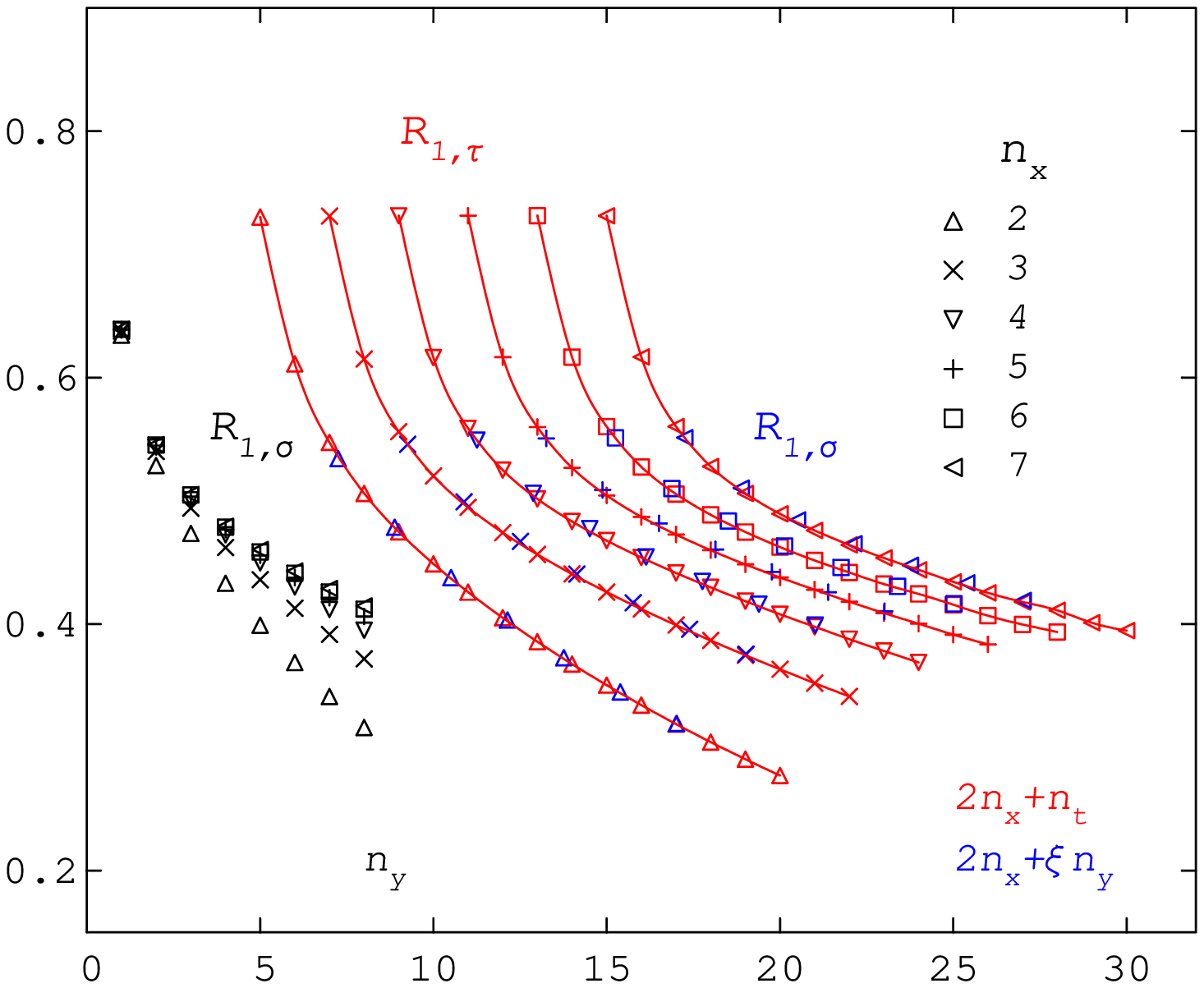, width=100mm,height=83mm}
\end{center}
\caption{The ratios $R_{1,\tau}(n_x,n_t)$ at $\beta=6.3$ and $\gamma=1.5$
for fixed $n_x=2,...,7$ plotted 
vs. $2n_x+n_t$ and connected by splines. The ratios 
$R_{1,\sigma}(n_x,n_y)$ are once plotted vs. $n_y$ and also shifted, that is
vs. $2n_x+\xi n_y$, with $\xi=1.63\,$.} 
\label{fig:rat1}
\begin{center}
   \epsfig{bbllx=94,bblly=264,bburx=483,bbury=587,
       file=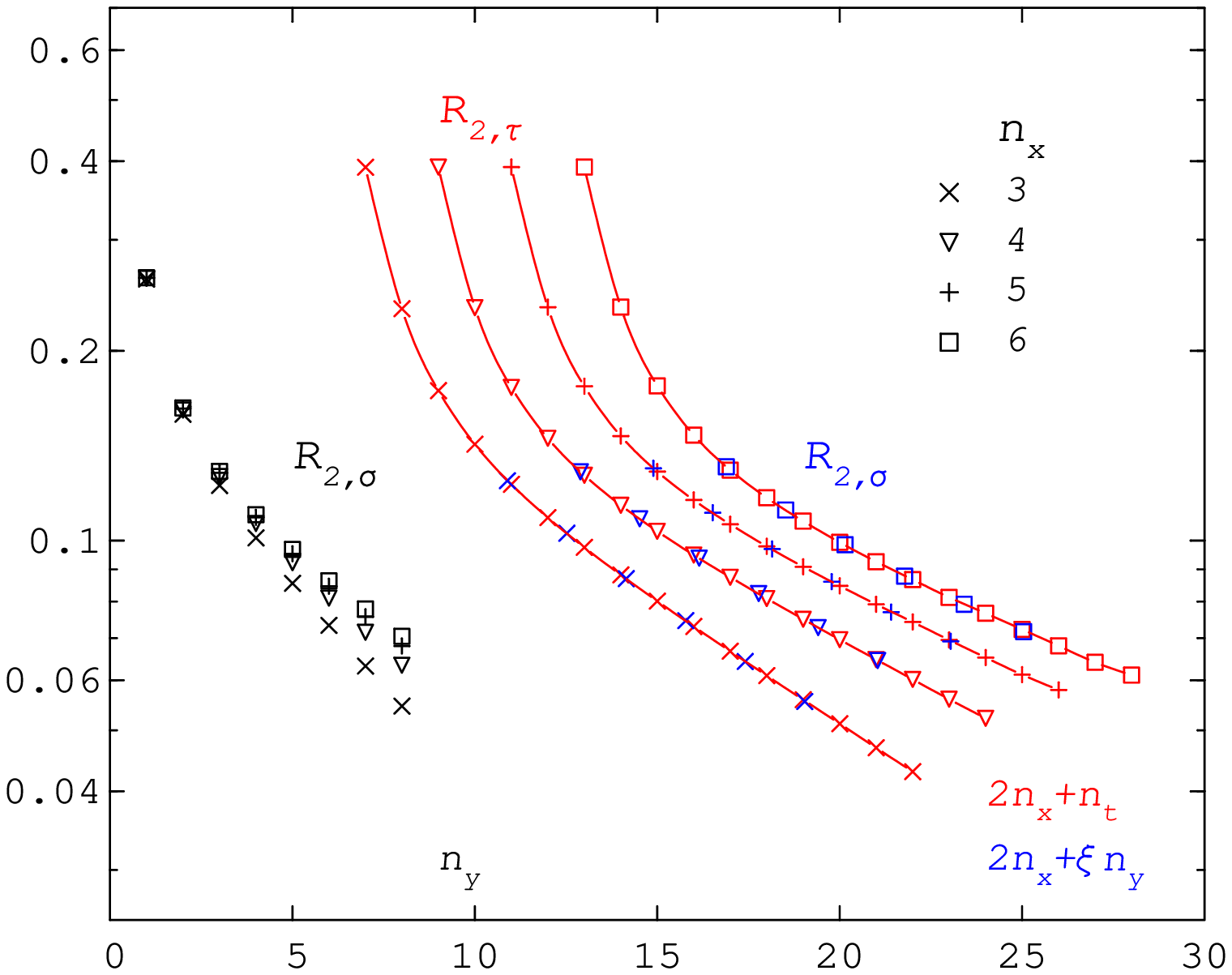, width=100mm,height=83mm}
\end{center}
\caption{The ratios $R_{2,\tau}(n_x,n_t)$ at $\beta=6.3$ and $\gamma=1.5$
for fixed $n_x=3,...,6$ plotted 
vs. $2n_x+n_t$ and connected by splines. The ratios 
$R_{2,\sigma}(n_x,n_y)$ are once plotted vs. $n_y$ and also shifted, that is
vs. $2n_x+\xi n_y$, with $\xi=1.63\,$.} 
\label{fig:rat2}
\end{figure}
\newpage
\n
by a factor $k$ such that the sum of the squared deviation 
($\chi^2$) of the shifted $R_{\sigma}-$values from the respective  
$R_{\tau}-$interpolations becomes minimal. For $\gamma<1$ the roles of  
$R_{\sigma}$ and $R_{\tau}$ are interchanged. The fitting is done for all 
ratios with $n_x\ge m_x$ and $n_y\ge m_y$ at the same time, with suitably chosen 
minimal values $m_x$ and $m_y$. In so far our procedure differs from the
matching prescription of Klassen \cite{Klass}, who matches single ratios and 
looks for a possible plateau of shift values. In Figs. 
\ref{fig:rat1} and \ref{fig:rat2}
we show matching examples at $\beta=6.3$ and $\gamma=1.5$ for the ratios    
$R_1$ and $R_2\,$. The matching in Fig. \ref{fig:rat1} is optimized for
$m_x=m_y=2\,$, in Fig. \ref{fig:rat2} for $m_x=m_y=3\,$. Both lead to the same
value $\xi=1.63$ for the anisotropy. For $R_{1}$ we find $k=1.01$ 
and $k=1.02$ for $R_2\,$, that is $k$ is always very close to one.
One may as well perform fits, as Klassen\cite{Klass} does, with fixed
$k=1$. The corresponding $\chi^2$ is then considerably larger, 
the value for $\xi$ increases slightly. We shall come back to this 
point again.

\begin{figure}[tb]
\begin{center}
   \epsfig{bbllx=62,bblly=264,bburx=516,bbury=587,
        file=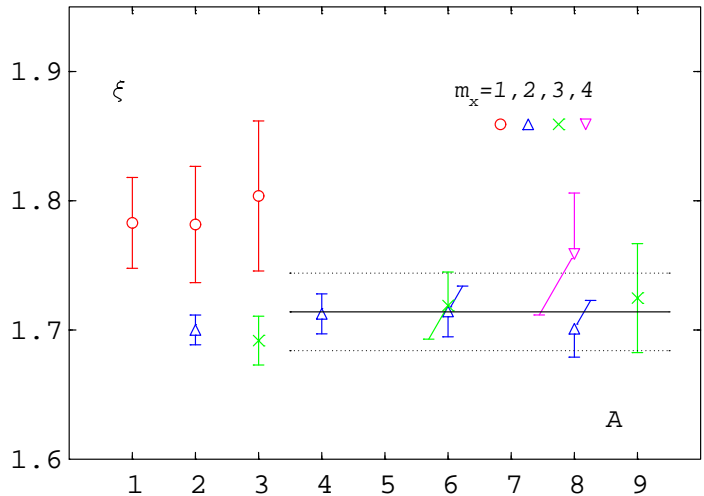,width=84mm,height=60mm}
\end{center}
\caption{The result for $\xi$ from the matching of $R_1$ at $\beta=5.7$
and $\gamma=1.5$ as a function of $A=m_xm_y\,$. In the matching all ratios 
with $n_x\ge m_x$ and $n_y\ge m_y$ were taken into account. The lines
show the final result with an error band. }
\label{fig:err57}
\end{figure}

As has been mentioned already, we have included only ratios in the matching
process which are built from Wilson loops with at least minimal extensions 
$m_x$ and
$m_y\,$. This has been done for two reasons. Once, $x_1$ should anyhow not
be too small\break (see eq. (\ref{Wilson})) and second, Wilson loops 
with a single
link on one side cannot be link integrated and are therefore less accurate.
That is why we disregard ratios with $n_x=1$ and/or $n_y=1$ in the final 
analysis. The influence of the chosen minimal values on the matching 
result for $\xi$ is demonstrated in Fig. \ref{fig:err57} for $R_1$ 
at $\beta=5.7$ and $\gamma=1.5\,$. We observe, that apart from the $m_x=1$
data all other measurements for $\xi$ are consistent with each other.  
 

\subsection{Results for \bf {$c_{\sigma,\tau}$ from matching}}

\begin{figure}[htb]
\begin{center}
   \epsfig{bbllx=94,bblly=264,bburx=483,bbury=587,
       file=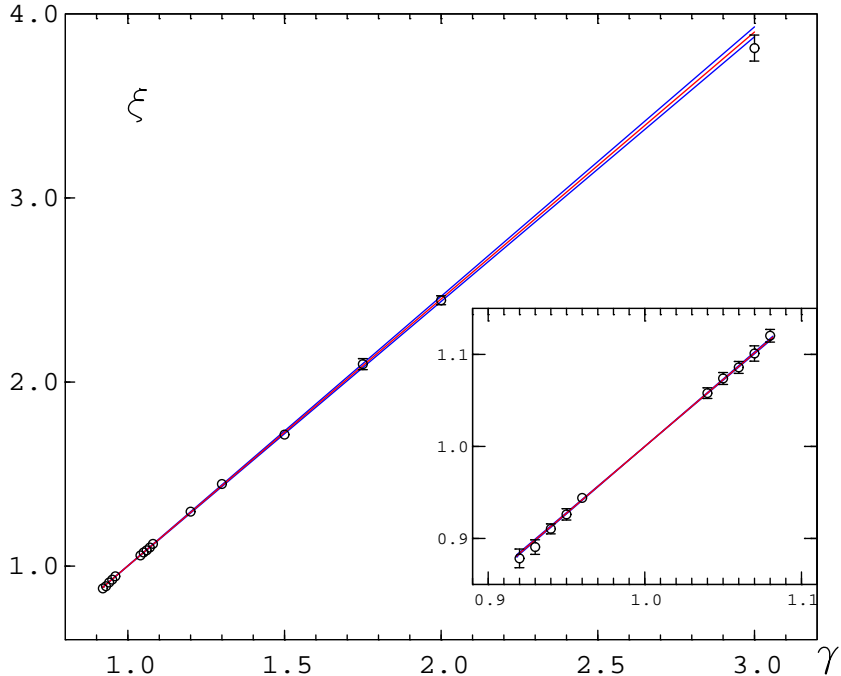, width=100mm,height=83mm}
\end{center}
\caption{The anisotropy $\xi(\gamma)$ at $\beta=5.7$ as determined from
the ratio $R_1$ with $m_x=2,m_y=3$. The red lines are from a linear fit   
in $\gamma$, the blue lines show the error band
.} 
\label{fig:xi}
\end{figure}

\n In order to obtain the anisotropy coefficients we measure now the
function $\xi(\gamma)$ in the neighbourhood of $\gamma=1$ with the aim of
determining the derivative $(\partial\xi/\partial\gamma)_{\xi=1}\,$.
In principle one would choose $\gamma$ rather close to one. However,
the difference between spatial and temporal Wilson loops becomes then 
very small as well and the matching will be inexact. Fortunately, it turns
out, that $\xi$ is linear in $\gamma$ in a wide range around $\gamma=1$
at all $\beta-$values. A successful strategy is therefore to measure
the Wilson loops with high precision at not too many $\gamma-$values to 
determine the slope of $\xi(\gamma)\,$. In Fig. \ref{fig:xi} we show as an
example the function
$\xi(\gamma)$ for the Wilson action at $\beta=5.7$ as obtained 
from the matching of $R_1$ with $m_x=2,m_y=3$, together with a linear fit
to the data points. 

As in the last section, we use now the $\beta-$function of ref. \cite{Boyd96}
to deduce the anisotropy coefficients from the measured derivative
$(\partial\xi/\partial\gamma)_{\xi=1}\,$. The corresponding

\begin{figure}[p]
\begin{center}
  \epsfig{bbllx=102,bblly=264,bburx=521,bbury=587,
          file=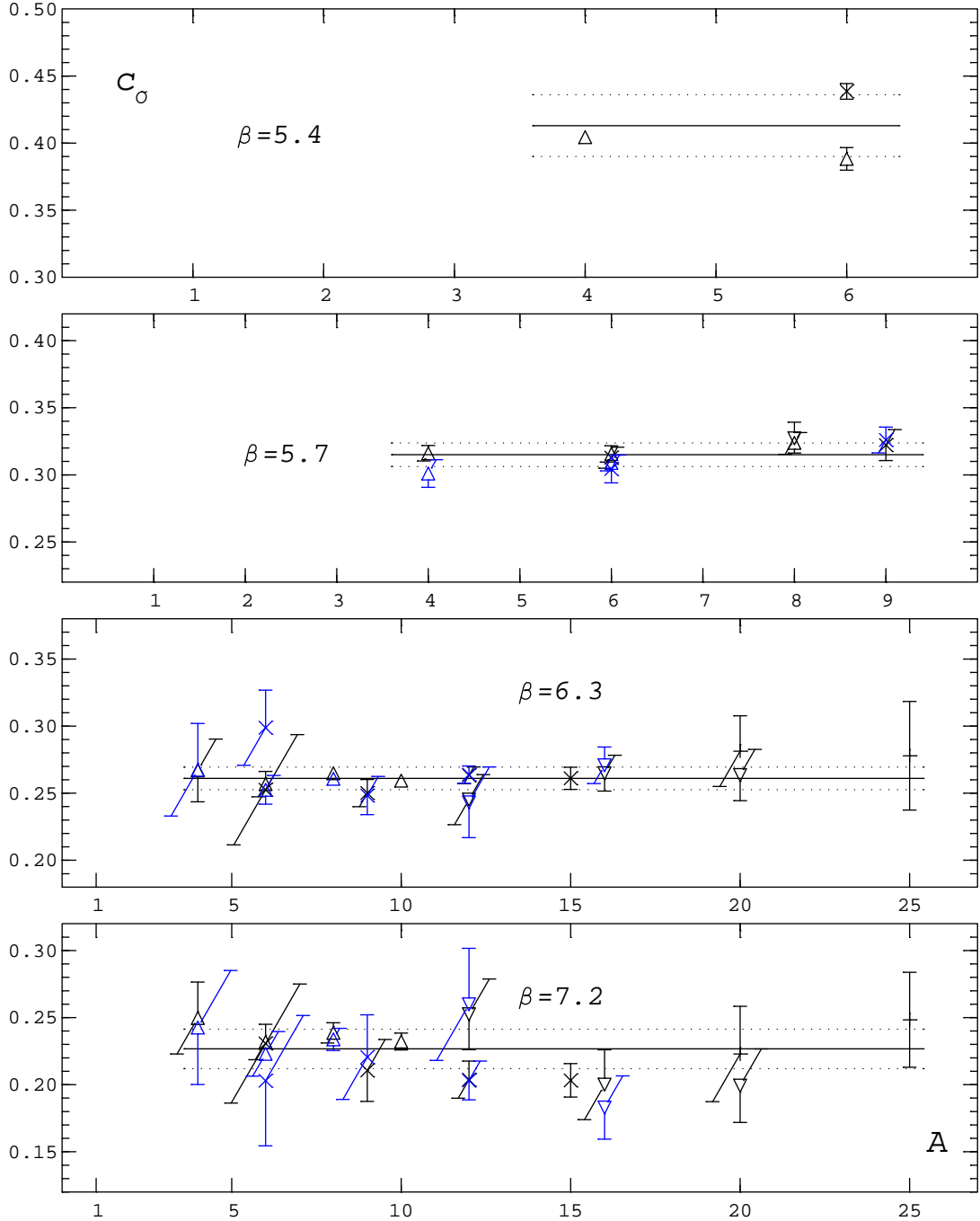,angle=-90,width=13.5cm}
\end{center}
  \caption{The anisotropy coefficient $c_{\sigma}$ of the Wilson action
  from the matching of 
   $R_1$ (black) and $R_2$ (blue points) as a function of $A=m_xm_y$ for
   $\beta=5.4,\,5.7,\,6.3,\,$ and $7.2\,$. Points with $m_x=2,3,4,5$ are denoted
   by $\bigtriangleup,\times,\bigtriangledown,+\,$, respectively.} 
    \label{fig:4ww}

\end{figure}

\n results for 
$c_{\sigma}$ are shown in Fig. \ref{fig:4ww} as a function of the minimal 
area $A=m_xm_y$ of the Wilson loops, which were included in the ratio-matching.
The data for $\beta=5.4$ are somewhat problematic in several respects and 
can only serve as an indication for the true value. As demonstrated already
in Fig. \ref{fig:err} too large Wilson loops cannot be included in the matching
for $\beta=5.4$, because their errors are too large. The second handicap is 
the unknown or ambiguous $\beta-$function in the strong coupling region.
We have therefore just taken the same value for $c_{\sigma}+c_{\tau}$ at 
$\beta=5.4$ and $5.7\,$. At $\beta \ge 5.7$ we find that the results are 
essentially independent on the minimal area, 
\begin{table}
\begin{center}
  \begin{tabular}{|c|cccc|}
    \hline
    $\beta$ & 5.4       & 5.7       & 6.3       & 7.2 \\ \hline
    $c_{\sigma}$   &$\,\,$0.413(23) &$\,\,$0.315(09) &
    $\,\,$0.261(08) &$\,\,$0.227(15) \\  
    $c_{\tau}$   & -0.374(23) & -0.276(09) & -0.202(08) & -0.159(15)
                 \\  \hline\hline
    $c_{\sigma}(k=1)$ &$\,\,$0.397(17) &$\,\,$0.348(06) &
    $\,\,$0.296(06) &$\,\,$0.261(12) \\ 
    $c_{\tau}(k=1)$   & -0.358(17) & -0.309(06) &
     -0.237(06) & -0.193(12) \\ \hline
  \end{tabular}
\end{center}
\caption{The matching results for $c_{\sigma}$ and $c_{\tau}$ for the Wilson
action .}
\label{tab:csct}
\end{table}
\begin{figure}[b]
\begin{center}
  \epsfig{bbllx=62,bblly=355,bburx=516,bbury=487,
          file=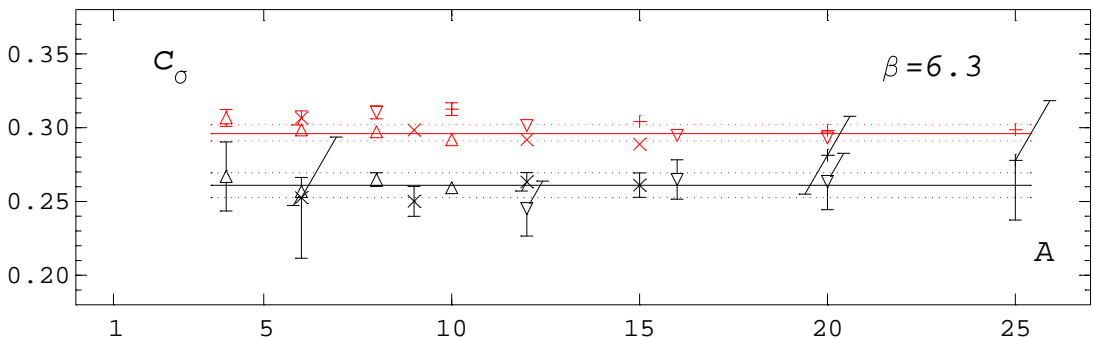,width=13.5cm}
\end{center}
  \caption{Comparison of the  anisotropy coefficient $c_{\sigma}$
   of the Wilson action from the matching of 
   $R_1$ with $k\ne 1$ (black) and $k=1$ (red points) as a function 
   of $A=m_xm_y$ for
   $\beta=6.3\,$. Points with $m_x=2,3,4,5$ are denoted
   by $\bigtriangleup,\times,\bigtriangledown,+\,$, respectively.} 
    \label{fig:comc}
\end{figure}
at $\beta=7.2$ the errors 
are larger because of the lower statistics of the measurements. The final 
results for $c_{\sigma}$ and $c_{\tau}$ are given in Table \ref{tab:csct}$\,$. 
We obtained them by fitting the single values to a constant with the 
$\chi^2-$method. Their average errors were estimated from the error estimate
of the constant times the square root of the number $N$ of fitpoints. The error 
estimate was rescaled whenever $\chi^2/(N-1)>2\,$. This was only necessary for
$\beta=5.4\,$. In order to be able to compare our results to those of Klassen
\cite{Klass}, we have repeated the matching analysis under the assumption 
$k=1\,$. Though the single matching results seem to have smaller errors then,
a constant $\chi^2-$fit as before yields always a $\chi^2$ per degree of
freedom which is larger than 2. The inclusion of small Wilson loops in the 
matching seems to be more crucial here and also to lead to higher values for
$c_{\sigma}\,$. This is observed in Fig. \ref{fig:comc}, where we compare the 
two methods at $\beta=6.3\,$. 
\begin{figure}[htb]
\begin{center}
   \epsfig{bbllx=94,bblly=264,bburx=483,bbury=587,
       file=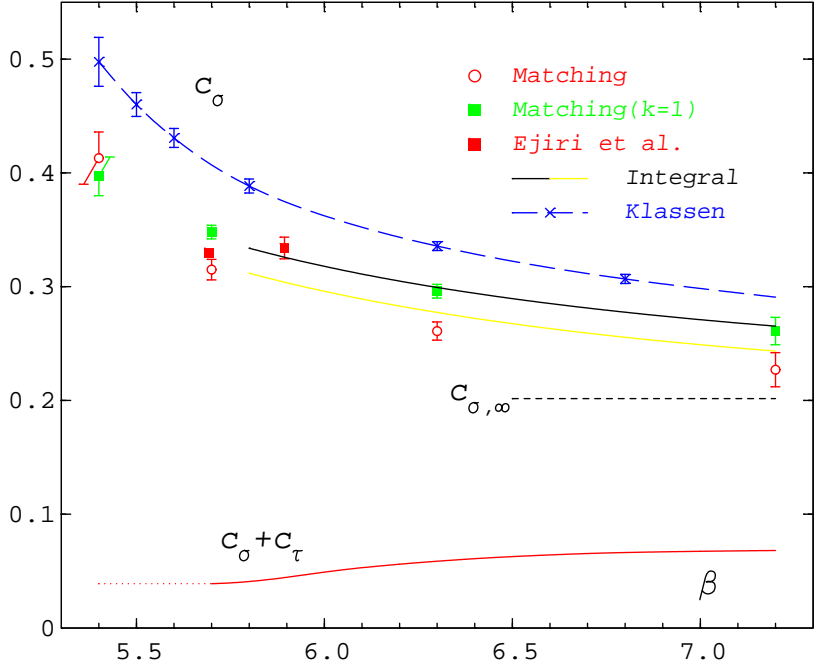, width=100mm,height=83mm}
\end{center}
\caption{Comparison of all available data for 
the  anisotropy coefficient $c_{\sigma}$ of the Wilson action.
Further details are explained in the text.} 

\label{fig:coma}
\end{figure}

Finally, we have gathered in Fig. \ref{fig:coma} all available results
for the Wilson action $c_{\sigma}\,$: those which we obtained from the
integral method are represented by the fit, eq.(\ref{pade}), and the fit
showing the possible influence of $p/T^4(\beta_0)\,$; our matching
results for $k\ne1$ and $k=1\,$; the $SU(3)$ data of Ejiri et al.\cite{Ejiri}
and those of Klassen\cite{Klass}, both for our $\beta-$function. We observe
consistent behaviour of the integral and matching results for $k=1\,$,
those for $k\ne1$ are still within the error bars (see Fig. \ref{fig:cs3b}).
There is also full agreement with the data of Ejiri et al.~. Only the results 
of Klassen are definitely higher and insofar incompatible with all other 
measurements of $c_{\sigma}\,$. 


\section{Anisotropy coefficients for improved actions}


Up to now we have discussed in detail the calculation of the anisotropy
coefficients for the Wilson action only. In the following we consider the
more general family of actions defined by Wilson loops of size up to 4 in the 
($\mu,\nu$)-plane
\begin{eqnarray}
  \label{sqs}
S_{\mu\nu}\!\!\!\!&=&\!\!\!\!
     a_{1,1}\left(1-\frac{1}{N_c}\re\tr\plaq\right) \nn\\
&&\!\!\!\!+ a_{1,2} \left(1-\frac{1}{2N_c}\re\tr\left(\loOp +\lOop\right)\right)
     +a_{2,2} \left(1-\frac{1}{N_c}\re\tr\loopt\right).
\end{eqnarray}
Here, the parameters $a_{11},~a_{12}$ and $a_{22}$ are constrained by the 
equation
\beqn
a_{11} + 4a_{12} + 16a_{22} = 1~,
\eqn
which ensures the correct continuum limit. The action is  $O(a^2)$ improved,
if additionally
\beqn
 a_{12} = {2 \over 3} - {a_{11} \over 2}~,\quad 
 a_{22} = {1 \over 16}\left(a_{11} -{5 \over 3}\right)~.
\eqn
Still, one parameter, for example $a_{11}$, is free. For $a_{11}=5/3$ we
obtain the Symanzik action ($a_{12}=-1/16,~a_{22}=0$), for $a_{11}=4/3$
the $(2\times 2)-$action ($a_{12}=0,~a_{22}=-1/48$). 
 Alternatively, one may require that the propagator can be diagonalized. 
This facilitates the analytic calculation of the anisotropy coefficients
in the weak coupling limit. In that case we have the condition
\beqn
 a_{11} = {1 \over (1+4z)^2}~, \quad  a_{12} = 2z a_{11}~,\quad
 a_{22} = z^2 a_{11}~,
\eqn
with the free parameter $z$. One can combine both objectives for a special 
set of parameters ($z=-1/16$)
\beqn
 a_{11} = {16 \over 9}~, \quad  a_{12} = -{2 \over 9}~,\quad
 a_{22} = {1 \over 144}~.
\eqn
The corresponding action was introduced by Garc\'\i a P\'erez et al.
\cite{Garca} under the name Square Symanzik action. In ref. \cite{Garcb}
the asymptotic $(\beta\rightarrow\infty)$ anisotropy coefficients 
of this action have actually been calculated. For $SU(3)$ one obtains
\beqn
c_{\sigma}(0)=0.09829281~,\quad c_{\tau}(0)=-0.03008298~.
\label{cinf}
\eqn
It is therefore obvious to investigate the properties of this action further
and in particular to determine its anisotropy coefficients non-perturbatively.
For comparison we 
have as well considered the somewhat simpler $(2\times 2)-$action, which is
improved, but its perturbative anisotropy coefficients are unknown. 

In order to apply the integral method we need again the high temperature 
limits of energy density and pressure. For the Square Symanzik action we
have derived the expansions corresponding to eqs. (\ref{RSW}) and 
(\ref{RIW})
\eqa
\!\!R_D(N_{\tau})\!\!\! &=&\!\!\!1
-{40 \over 231}\left( {\pi\over N_{\tau}}\right)^6
-{176896 \over 3869775} \left( {\pi\over N_{\tau}}\right)^8 +
O\left(N_{\tau}^{-10}\right)\\ \nonumber
\!\!\! & \approx&\!\!\!1-0.1732 \left( {\pi\over N_{\tau}}\right)^6
 -0.0457\left( {\pi\over N_{\tau}}\right)^8 +
O\left( N_{\tau}^{-10}\right)~;\\
\!\!R_I(N_{\tau})\!\!\! &=&\!\!\!1
-{8 \over 105}\left( {\pi\over N_{\tau}}\right)^4
-{80 \over 693}\left( {\pi\over N_{\tau}}\right)^6
-{420128 \over 4729725} \left( {\pi\over N_{\tau}}\right)^8 +
O\left(N_{\tau}^{-10}\right)\\ \nonumber
 & \approx&\!\!\! 1-0.076 \left( {\pi\over N_{\tau}}\right)^4
-0.115 \left( {\pi\over N_{\tau}}\right)^6
 -0.089\left( {\pi\over N_{\tau}}\right)^8 +
O\left( N_{\tau}^{-10}\right)~.
\ena

\n By chance, the ratio $R_D(N_{\tau})$ is even $O(a^4)$ improved. Since we
intend to make use of Monte Carlo simulations on $\nt=4$ lattices we list in 
Table \ref{tab:Rimp} the respective numerical results for the complete
$1/N_{\tau}-$expansions of the ratios. The numbers for the 
$(2\times 2)-$action have been taken from ref. \cite{Bein96}.

\begin{table}
\begin{center}
\begin{tabular}{|c|c|c|c|}
\hline
Action & $R_D$ & $R_I$ & $r=R_D/R_I$  \\
\hline
$(2\times 2)$  & 1.088 & 0.99 & 1.099  \\
Square Sym.    & 0.957 & 0.91 & 1.052  \\
\hline
\end{tabular}
\end{center}
\caption{ Derivative ($R_D$) and integral method ($R_I$) ratios at 
$N_{\tau}=4$ for the $(2\times 2)$ and the Square Symanzik actions.}
\label{tab:Rimp}
\end{table}
 

\subsection{Results for the ${\bf (2\times 2)-}$action}


The thermodynamics of the $SU(3)~(2\times 2)-$action has already been 
investigated in detail by simulations on $24^3\times 4$ and $24^4$ lattices in 
ref. \cite{Bein96}. We can therefore immediately apply the integral method
to compute the anisotropy coefficients using the corresponding plaquette
expectation values and the $\beta-$function. For \nt=4 the critical coupling
was found to be $\beta_c=4.3995(2)\,$. The behaviour of $c_{\sigma}$ is
similar to that shown in Fig. \ref{fig:cs3a} for the Wilson action. With 
increasing $\beta$ $c_{\sigma}$ is slowly decreasing. In the neighbourhood 
of the critical point the integral method leads again to numerical difficulties. 
\begin{table}[t]
\begin{center}
  \begin{tabular}{|c|ccc|}
    \hline
    $\beta$ & 4.4       & 5.0       & 5.9    \\ \hline
    $c_{\sigma}$   &$\,\,$0.226(08) &$\,\,$0.157(16) &
    $\,\,$0.115(16) \\  
    $c_{\tau}$   & -0.190(08) & -0.096(16) & -0.067(16)
                 \\  \hline\hline
    $c_{\sigma}(k=1)$ &$\,\,$0.236(03) &$\,\,$0.195(07) &
    $\,\,$0.158(13) \\ 
    $c_{\tau}(k=1)$   & -0.200(03) & -0.134(07) &
     -0.110(13) \\ \hline
  \end{tabular}
\end{center}
\caption{The matching results for $c_{\sigma}$ and $c_{\tau}$ for the 
$(2\times 2)-$action.}
\label{tab:cst22}
\end{table}
\begin{figure}[b]
\begin{center}
   \epsfig{bbllx=94,bblly=264,bburx=483,bbury=587,
       file=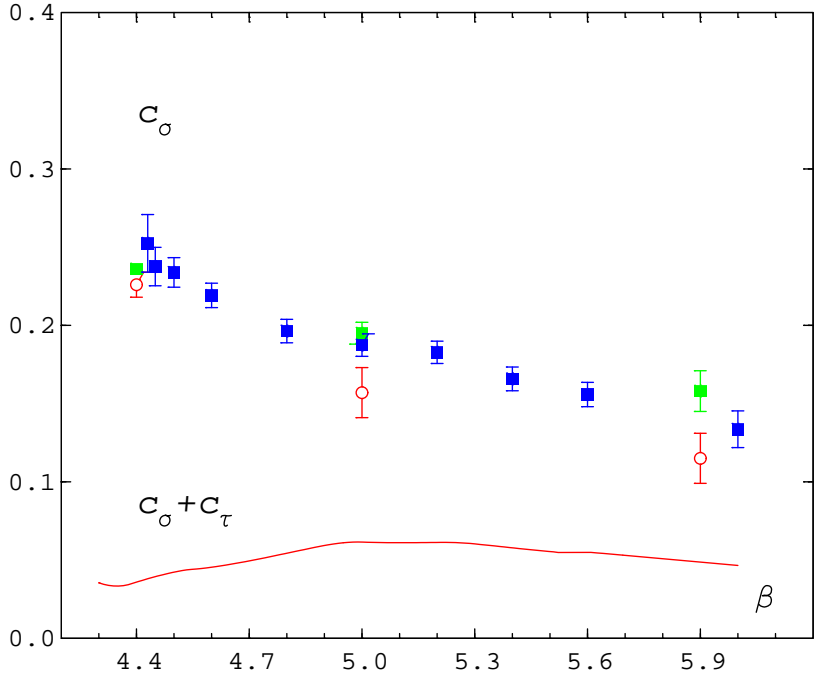, width=100mm,height=83mm}
\end{center}
\caption{The anisotropy coefficient $c_{\sigma}$ of the $(2\times 2)-$action
from the integral (blue squares) and matching methods (circles). Results from 
fixed $k=1$ matching fits are shown as green squares.} 
\label{fig:cs22}
\end{figure}
In Fig. 
\ref{fig:cs22} we compare the integral method results with matching results,
which were obtained on anisotropic $24^4$ lattices with 2000-4000 Wilson loop
measurements for $\beta=4.4$ and 5.0 and 600 for $\beta=5.9\,$. We note, that
due to the inclusion of $(2\times 2)-$Wilson loops in the action, less links
can be integrated to increase the accuracy of the measurements. In Table
\ref{tab:cst22} we present the values for $c_{\sigma}$ and $c_{\tau}$ from
the matching procedure. Again we observe, that the matching with fixed $k=1$
leads to somewhat higher values. We find agreement between the different 
methods, but like for the case of the Wilson action only after taking into 
account the cut-off correction factor $r$ from Table \ref{tab:Rimp}.
 

\subsection{Results for the Square Symanzik action}

\begin{figure}[tb]
\begin{center}
   \epsfig{bbllx=94,bblly=264,bburx=483,bbury=587,
       file=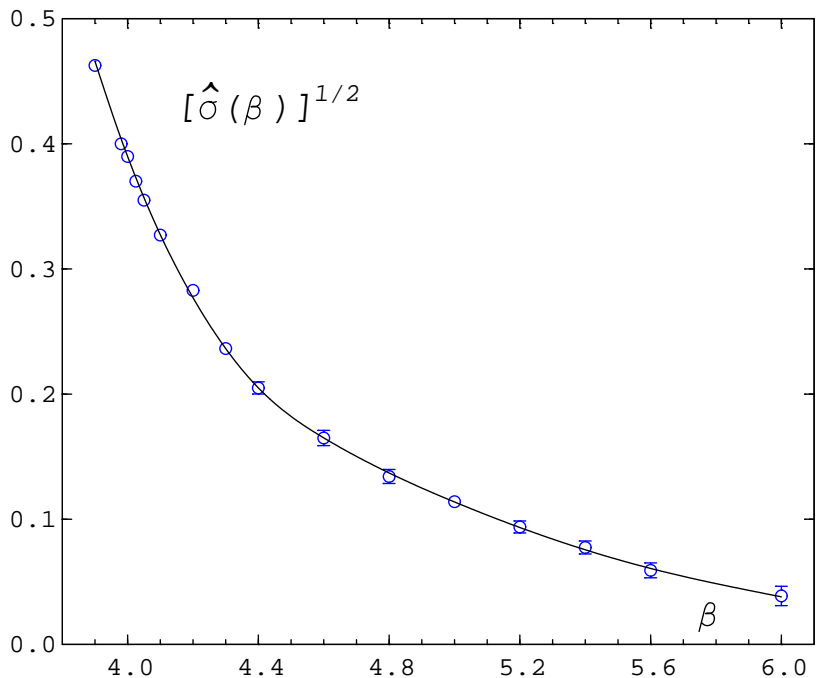, width=100mm,height=83mm}
\end{center}
\caption{The square root of the string tension, $\sqrt{\hat \sigma(\beta)}$
for the $SU(3)$ Square Symanzik action. The line is a fit to the data.}
\label{fig:sigma}
\end{figure}

\begin{figure}
\begin{center}
   \epsfig{bbllx=94,bblly=264,bburx=483,bbury=587,
       file=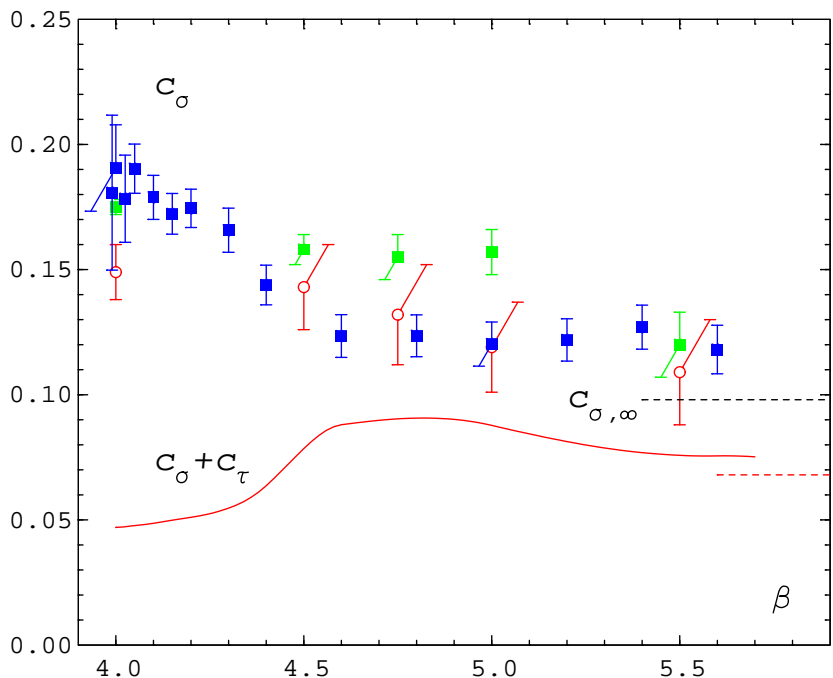, width=100mm,height=83mm}
\end{center}
\caption{The anisotropy coefficient $c_{\sigma}$ of the Square Symanzik action
from the integral (blue squares) and matching methods (circles). Results from 
fixed $k=1$ matching fits are shown as green squares. The dashed black line 
shows $c_{\sigma,\infty}\,$.} 
\label{fig:csqs}
\end{figure}

In order to determine the $\beta-$function for this action we have performed 
simulations on $24^4$ lattices and measured the plaquettes and the string 
tension as in ref. \cite{Lego}. The lattice result $\hat \sigma(\beta)$ and
the physical string tension $\sigma$ are related by $\hat \sigma(\beta)=
\sigma a^2\,$. The $\beta-$function is then derived from
\beqn
a{dg^{-2} \over da} = {a \over 2N_c}{d\beta \over da} = 
{\sqrt{\hat\sigma} \over 2N_c} \left( {d\sqrt{\hat\sigma(\beta )} \over
d\beta} \right)^{-1}~. 
\eqn
In Fig. \ref{fig:sigma} we show our measured values for 
$\sqrt{\hat \sigma(\beta)}$ together with a fit from which we have deduced 
the numerical values of the $\beta-$function. On a $24^3\times 4$ lattice
we have then measured the generalized plaquettes $P_{\sigma}$ and $P_{\tau}$
as well as the Polyakov loop and its susceptibility in the range $3.9\le \beta
\le 6.0\,$. On the average we took 4000-8000, close to the critical 
point 18000 measurements. From the peak of the susceptibility we estimate
the following critical coupling for $\nt=4$ 
\begin{eqnarray}
  \label{bc-erg}
  \beta_c=3.9820\,(  
  \renewcommand{\arraystretch}{0.5}
  \begin{array}[c]{c}
    \scriptstyle +8\\
    \scriptstyle -5
  \end{array}
  \renewcommand{\arraystretch}{2}
   )~.
\end{eqnarray}
\n As a side product of our calculations we obtain the ratio
\beqn
{T_c \over \sqrt{\sigma}} = {1 \over \nt\sqrt{\hat \sigma(\beta_c)}} = 
0.643(4)~, 
\eqn
which is in agreement with results found for other improved actions \cite{Bklp}.
Using the above mentioned data and $r$ from Table \ref{tab:Rimp} we have
again applied the integral method to find the anisotropy coefficient 
$c_{\sigma}\,$. It is shown in Fig. \ref{fig:csqs} together with the 
asymptotic value $c_{\sigma,\infty}\,$, eq. (\ref{cinf}), and the results of 
our matching analysis. The latter are also listed in Table \ref{tab:cstsqs}.
They were obtained from Wilson loop measurements on $24^4$ lattices. Like
for the $(2 \times 2)-$action less link integrations than in the Wilson 
action case can be done. Comparing the different results in Fig. \ref{fig:csqs}
we find again the same general $\beta-$dependence as for the other actions.
Also, a sizeable difference between the different matching options for $k$
is found: the $k=1$ results show an even stronger influence of the 
Wilson loop area  
as before, the $k\ne 1$ results agree nicely with the integral method data.
We note that $c_{\sigma}$ in the Square Symanzik case is about 
half as big as in the Wilson case. 

\begin{table}
\begin{center}
  \begin{tabular}{|c|ccccc|}
    \hline
    $\beta$ & 4.0       & 4.5       & 4.75      & 5.0   & 5.5 \\ 
    $N_{meas}$ & 2000 & 800-1000& 700 & 500-1500 & 200\\ \hline
    $c_{\sigma}$   &$\,\,$0.149(11) &$\,\,$0.143(17) &
    $\,\,$0.132(20) &$\,\,$0.119(18) &$\,\,$0.109(21) \\  
    $c_{\tau}$   & -0.102(11) & -0.065(17) & -0.042(20) & -0.031(18)
                 & -0.033(21) \\  \hline\hline
    $c_{\sigma}(k=1)$ &$\,\,$0.175(03) &$\,\,$0.158(06) &
    $\,\,$0.155(09) &$\,\,$0.157(09) &$\,\,$0.120(13) \\ 
    $c_{\tau}(k=1)$   & -0.128(03) & -0.080(06) &
     -0.065(09) & -0.069(09) & -0.044(13) \\ \hline
  \end{tabular}
\end{center}
\caption{The matching results for $c_{\sigma}$ and $c_{\tau}$ for the Square
Symanzik action. $N_{meas}$ is the number of Wilson loop measurements .}
\label{tab:cstsqs}
\end{table}

\section{Summary and conclusions}

\n One of the major objectives of our paper was the verification of the
equivalence of the integral and derivative methods for the calculation of 
thermodynamic quantities like energy density and pressure. The major obstacle
in the comparison of the methods is the cut-off dependence on finite lattices.
In the high temperature limit the deviations of energy density and pressure 
from their respective continuum limits due to the cut-off effects can be 
calculated. They manifest themselves as a dependence on $1/\nt^2$ in the
thermodynamic limit, and they are different for the two methods.

 The anisotropy coefficients as such must be finite size independent. Their 
calculation allows then for a check on the internal consistency of the 
integral method and of its equivalence to the derivative method. Moreover, its
non-perturbative behaviour is of importance for the approach of thermodynamic
quantities to the continuum limit.

 By taking into account the major cut-off effects through the
correction factor $r$ we have shown for the Wilson action that one obtains 
similar results for \nt=4, 6 and 8 which agree, inside error bars, also with
the matching results and with the results of Ejiri et al.\cite{Ejiri}. The 
$c_{\sigma}-$data of Klassen\cite{Klass} are however definitely higher.
This may be connected to the difference in our matching methods, the 
accuracy of the data and the lattice sizes used. We found it essential for
the application of the matching procedure, that link integration of the Wilson 
loops was carried out, wherever this was possible, and that ratios instead of 
Wilson loops directly were used for the matching. Thus we avoided 
perimeter and corner dependencies. Remaining effects of this kind are probably 
responsible for the difference of $k=1$ to $k\ne 1$ matching fits. The fact that 
$\xi(\gamma)$ was linear near $\gamma=1$ helped in the determination of the 
derivative $(\partial\xi/\partial\gamma)_{\xi=1}$.

 We have repeated our analysis for two improved actions, the $(2\times 2)$
and the Square Symanzik action, and find similar behaviours of the 
corresponding anisotropy coefficients. For all investigated actions the 
non-perturbative $c_{\sigma}$ decrease with increasing $\beta$ and approach
their respective weak coupling limits from above; the results from the matching 
and integral methods are in all cases compatible with each other.

In addition we determined for the Square Symanzik action the deconfinement
transition point $\beta_c$ for $\nt=4\,$, the ratio $T_c/\sqrt{\sigma}$ and
the $\beta-$function in terms of the string tension. As an analytic result
we derived the thermodynamic high temperature limits for this action.


\bn {\Large \bf Acknowledgements}

\bn
We thank Maria Garc\'\i a P\'erez for providing us with details of her 
work in refs. \cite{Garca} and \cite{Garcb}. The help of Burkhard Sturm
(alias B. Beinlich)
in the calculation of the determinant of the inverse propagator and the high
temperature limits is gratefully acknowledged. The work was partly
supported by the Deutsche Forschungsgemeinschaft under grants Pe 340/3-3
and Ka 1198/4-1, and the European Commission under the
TMR-Network "Finite Temperature Phase Transitions in Particle Physics",
ERBFMRX-CT-97-0122.

\end{document}